\documentclass{elsart}
\usepackage{graphicx}
\usepackage{amssymb}
\renewcommand{\thefootnote}{\fnsymbol{footnote}}
\renewcommand{\thesection}{\arabic{section}.}
\renewcommand{\thesubsection}{\arabic{section}.\arabic{subsection}.}

\begin{document}
\def \ee {\varepsilon}
\def \ef {{\varepsilon^{(1)}}}
\def \es {{\varepsilon^{(2)}}}
\def \ezf {{\varepsilon_0^{(1)}}}
\def \ezs {{\varepsilon_0^{(2)}}}
\def \rf {{r_0^{(1)}}}
\def \rs {{r_0^{(2)}}}

\thispagestyle{empty}
\begin{frontmatter}
\title{
Analytic approach to the thermal Casimir force between metal and dielectric}

\author[a]{B.~Geyer,}
\author[a,b]{G.~L.~Klimchitskaya,}
\author[a,c]{V.~M.~Mostepanenko\corauthref{1}
}

\address[a]{
Center of Theoretical Studies and Institute for Theoretical Physics,\\
Leipzig University, Augustusplatz 10/11, D-04109 Leipzig, Germany 
}
\address[b]{
North-West Technical University, Millionnaya St. 5,
St.Petersburg, 191065, Russia}
\address[c]{
Noncommercial Partnership  ``Scientific Instruments'', 
Tverskaya St. 11, Moscow, 103905, Russia}

\corauth[1]{Corresponding author. Fax: +49~341~9732548
\\E-mail address:
{Vladimir.Mostepanenko@itp.uni-leipzig.de}}

\begin{abstract}
The analytic asymptotic expressions for the Casimir free energy, 
pressure and entropy at low temperature in the configuration of 
one metal and one dielectric plate are obtained. For this purpose 
we develop the perturbation theory in a small parameter proportional 
to the product of the separation between the plates and the 
temperature. This is done using both the simplified model of an 
ideal metal and of a dielectric with constant dielectric permittivity
and for the realistic case of the metal and dielectric with 
frequency-dependent dielectric permittivities. The analytic 
expressions for all related physical quantities at high temperature 
are also provided. The obtained analytic results are compared with 
numerical computations and good agreement is found. We demonstrate 
for the first time that the Lifshitz theory, when applied to the 
configuration of metal-dielectric, satisfies the requirements of 
thermodynamics if the static dielectric permittivity of a dielectric 
plate is finite. If it is infinitely large, the Lifshitz formula
is shown to violate the Nernst heat theorem. The implications of 
these results for the thermal quantum field theory in Matsubara 
formulation and for the recent measurements of the Casimir force 
between metal and semiconductor surfaces are discussed.         
\end{abstract}
\end{frontmatter}

\maketitle

\section{Introduction}

The Casimir effect \cite{1} is the direct manifectation of zero-point
oscillations of quantized fields. It finds multidisciplinary applications
in quantum field theory, gravitation and cosmology, atomic physics,
condensed matter and, most recently, in nanotechnology (see, e.g., the
monographs \cite{2,3,4,5} and reviews \cite{6,7,8,9}). According to
Casimir's prediction, the existence of zero-point oscillations leads to 
the polarization of vacuum in quantization volumes restricted by material
boundaries and in spaces with non-Euclidean topology. This is accompanied
by forces acting on the boundary surfaces (the so called Casimir force).
The Casimir force acts between electrically neutral closely spaced
surfaces. It is a pure quantum phenomenon (there is no such a force in
the framework of classical electrodynamics) being the generalization
of the well known van der Waals force for the case of relatively large
separations where relativistic effects become essential.

The theoretical basis for the description of both the van der Waals and
Casimir forces is given by the Lifshitz theory \cite{10,11,12}. The
main formulas of the Lifshitz theory express the free energy and pressure
of the van der Waals and Casimir interaction between two plane parallel
plates as some functionals of the frequency-dependent dielectric 
permittivities of plate materials. These formulas can be derived in
many different theoretical schemes \cite{2,3,4,5,6,7,8,9,10,11,12}. In
particular, they were obtained in the framework of thermal quantum field
theory in the Matsubara formulation \cite{8}. During the last few years
the Lifshitz theory was successfully applied to the interpretation of
many measurements of the Casimir force between metal surfaces
\cite{13,14,15,16,17,18,19,20,21,22,23,24} and between metal and
semiconductor \cite{25,25a,25b,25c}.

A complicated problem of the Lifshitz theory is how to describe the
Casimir interaction between real metals at nonzero temperature. The
most convenient form of the Lifshitz formulas exploits the dielectric
permittivity along the imaginary frequency axis. The latter is obtained
from the tabulated optical data for the complex index of refraction
by means of the Kramers-Kronig relations. The available data are, however,
insufficient and must be extrapolated in some way to lower frequencies.
In this respect the contribution from the zero frequency is of most
concern. The point is that in Matsubara thermal field theory the
zero-frequency term becomes dominant at large separations (high
temperatures) whereas the contributions from all other Matsubara frequencies
being exponentially small. In \cite{26,27,28} the zero-frequency term
of the Lifshitz formula was obtained by using the dielectric function of
the Drude model. This results in a violation of the Nernst heat theorem in
the case of perfect crystal lattices \cite{29,30} and is in contradiction
with experiments at separations below 1$\mu$m \cite{22,23,24,31}. The
asymptotic value of the Casimir force at large separations predicted in
\cite{26,27,28} is equal to one half of the value predicted for
ideal metals, i.e., to one half of the so called classical limit \cite{32,33}.

Another approach \cite{34,35} uses the dielectric permittivity of the
plasma model to determine the zero-frequency term of the Lifshitz formula.
This approach was shown to be in agreement with thermodynamics \cite{29,30}
and consistent with experiment \cite{23}. It predicts the magnitudes of the 
Casimir force at short separations in qualitative agreement with the case
of ideal metals. At large separations the predicted force magnitude is
practically equal to that for ideal metals. Very similar results, which are
also in agreement with the requirements of thermodynamics and consistent
with experiment, are predicted by the surface impedance approach \cite{36,37}.
The controversies among different approaches to the thermal Casimir force
between metals are detailly discussed in \cite{28,30,37,38,39,40,41}.

Recently it was demonstrated \cite{42,43,43a} that even the traditional 
application of the Lifshitz formula to the case of two dielectric
semispaces presents problems. In \cite{42,43,43a} the analytic
asymptotic expressions for the Casimir free energy, pressure and entropy
at low temperatures (short separations) were found for two dielectrics.
It was shown that if the dielectric materials possess finite static
dielectric permittivities the theory is self-consistent and in agreement 
with thermodynamics. If, however, a nonzero dc conductivity of dielectrics 
is taken into account (any dielectric at nonzero temperature is
characterized by some nonzero dc conductivity which is many orders of
magnitude lower than for metals), this leads to a qualitative
enhancement of the Casimir force and a simultaneous violation of
the Nernst heat theorem. (Note that the dc conductivity of dielectrics
was taken into account in \cite{44} to explain the large observed
effect in noncontact friction \cite{45}.) In \cite{42,43,43a} the
phenomenological prescription was proposed 
that the dc conductivity of dielectrics is not
related to the Casimir interaction, and to avoid contradictions with
thermodynamics it should not be included in the
model of dielectric response.

The difficulties which were met in the application of the Lifshitz
theory to two metal and two dielectric plates attracted attention
to the case when one plate is metallic and another one dielectric.
This configuration was first considered in \cite{46}. It
presents the interesting opportunity to investigate the Casimir force
in the case when different plates are described by quite different
models of the dielectric response. In \cite{46}, however, only
the first leading terms in the low-temperature asymptotic expressions
for the Casimir free energy and entropy were obtained, and the
pressure was derived only in the dilute approximation. In the analytical
derivations in \cite{46} (see also the review \cite{43a})
it was supposed that the metallic plate
is made of ideal metal and the dielectric of the other plate is
described by the frequency independent dielectric permittivity. These
suppositions narrow the applicability of the obtained results. Also,
the role of the dc conductivity of a dielectric plate was not
investigated for plates with frequency-dependent dielectric
permittivities. 

In the present paper we develop the analytic approach to the thermal 
Casimir force acting between metal and dielectric
permitting to find several expansion terms in the asymptotic expressions
for all physical quantities at low temperature. 
This approach is applied not only
to the configuration of ideal metal and dielectric with frequency independent 
dielectric permittivity but also to real metal and dielectric described
by the dielectric permittivities depending on frequency. 
We pioneer in derivation of the low-temperature
asymptotic expressions for the Casimir free energy, entropy and
pressure between real metal and dielectric. 
The asymptotic behavior of all physical quantities at high
temperatures (large separations) is also provided. 
What is more, the obtained
representation for the Casimir free energy permits to find the low-temperature
behavior of the Casimir force acting between a metal sphere and a dielectric
plate (or, alternatively, dielectric sphere above a metal plate). This can be
done with the help of the proximity force theorem \cite{51}. 
The configuration of a
sphere above a plate is most topical in experiments on the measurement
of the Casimir force 
\cite{13,14,15,16,17,18,19,20,21,22,23,24,25,25a,25b,25c}.
(Note that for the experimental parameters the error introduced by the use
of the proximity force theorem was recently shown to be less than 0.1\%
\cite{47,48,49,50}.) 
Thus, our results will find immediate utility in experiment.
The analytic expressions for the Casimir interaction between metal and
dielectric at zero temperature are also found here for the first time. 
We determine the applicability region
of the obtained analytic formulas by compairing them with numerical 
computations using the tabulated optical data for 
metallic and dielectric materials. The fundamental
conclusion following from our results is that the Lifshitz theory, applied
to the configuration of a metal and a dielectric plate, is in agreement
with the Nernst heat theorem if the static dielectric permittivity of a
dielectric plate is finite. 
Note that this conclusion could not be achieved by using the numerical
computations which inevitably identify zero with all nonzero numbers in the
limits of a computational error.
If, however, the dc conductivity of a
dielectric plate is included in the model of dielectric response, we show
that the Nernst heat theorem is violated. This is in analogy to the same
conclusion in \cite{42} obtained for the configuration of two
dielectric plates and confirms our phenomenological prescription
that the dc conductivity is not related
to the van der Waals and Casimir forces and should not be included
in the model of dielectric response. Recently this prescription was
confirmed experimentally \cite{25c}.

The paper is organized as follows. In Section~2 we summarize the main formulas 
of the Lifshitz theory for the configuration of one plate made of metal
and another one made of dielectric. Section~3 is devoted to the simplified
model where the metal is an ideal one and dielectric is described by a
constant dielectric permittivity. In the framework of this model a
perturbation formalism applicable at low temperatures (short separations)
is developed. In Section~4 the realistic case is considered when the
dielectric permittivities of both metal and dielectric plates depend
on the frequency. The analytic asymptotic expressions for the free energy,
entropy and pressure of the Casimir interaction at both low and high 
temperatures are obtained. Section~5 contains the comparison between the 
analytical results and numerical computations using the tabulated optical 
data for plate materials. The application region of the derived asymptotic
expressions is determined. In Section~6 it is shown that the inclusion of the 
dc conductivity in the description of dielectric plate leads to a violation 
of the Nernst heat theorem. Section~7 contains our conclusions and discussion.

\section{Lifshitz formula in the configuration of metal and dielectric plates}

We consider two thick parallel plates (semispaces) at temperature $T$
in thermal equilibrium separated by the empty gap of width $a$. One plate
is made of metal with the dielectric permittivity $\varepsilon^M(\omega)$
and another of dielectric with permittivity $\varepsilon^D(\omega)$. The
free energy of the van der Waals and Casimir interaction between the plates 
per unit area is given by the Lifshitz formula \cite{10,11,12,42,46}
\begin{eqnarray}
&&
{\mathcal F}(a,T)=\frac{k_BT}{2\pi}\sum\limits_{l=0}^{\infty}
\left(1-\frac{1}{2}\delta_{0l}\right)\int_0^{\infty}
k_{\bot}dk_{\bot}
\label{eq1} \\
&&\phantom{aaaaa}
\times\left\{
\ln\left[1-r_{\|}^M(\xi_l,k_{\bot})r_{\|}^D(\xi_l,k_{\bot})e^{-2aq_l}
\right]\right.
\nonumber \\
&&\phantom{aaaaaaaa}\left.
+\ln\left[1-r_{\bot}^M(\xi_l,k_{\bot})r_{\bot}^D(\xi_l,k_{\bot})e^{-2aq_l}
\right]\right\}.
\nonumber
\end{eqnarray}
\noindent
Here the plates are perpendicular to the $z$ axis, 
$k_{\bot}=|{\mbox{\boldmath$k$}}_{\bot}|$
is the magnitude of the wave vector in the plane of plates,
$\xi_l=2\pi k_BTl/\hbar$ are the Matsubara frequencies, and
$k_B$ is the Boltzmann constant.
$r_{\|,\bot}^{M,D}$ are the reflection coefficients for metal ($M$)
and dielectric ($D$) plates for the two independent polarizations
of electromagnetic field calculated along the imaginary frequency
axis. Index $\|$ stands for the electric field parallel to the plane
formed by $\mbox{{\boldmath$k$}}_{\bot}$ and the $z$ axis
(transverse magnetic field), and index $\bot$ stands for the
electric field perpendicular to this plane (transverse electric field).
The explicit expressions for the reflection coefficients are
\cite{42,46}
\begin{equation}
r_{\|}^{M,D}(\xi_l,k_{\bot})=
\frac{\varepsilon_l^{M,D}q_l-k_l^{M,D}}{\varepsilon_l^{M,D}q_l+
k_l^{M,D}}, \qquad
r_{\bot}^{M,D}(\xi_l,k_{\bot})=
\frac{k_l^{M,D}-q_l}{k_l^{M,D}+q_l},
\label{eq2}
\end{equation}
\noindent
where
\begin{equation}
q_l=\sqrt{\frac{\xi_l^2}{c^2}+k_{\bot}^2}, \qquad
k_l^{M,D}=\sqrt{\varepsilon_l^{M,D}\frac{\xi_l^2}{c^2}+
k_{\bot}^2},
\label{eq3}
\end{equation}
\noindent
and
\begin{equation}
\varepsilon_l^{M,D}=\varepsilon^{M,D}(i\xi_l).
\label{eq4}
\end{equation}

The pressure of the van der Waals and Casimir interaction between
metal and dielectric (i.e., the force per unit area of plates) is
obtained from
\begin{equation}
P(a,T)=-\frac{\partial{\mathcal F}(a,T)}{\partial a}.
\label{eq5}
\end{equation}
\noindent
Using Eq.~(\ref{eq1}) we arrive at
\begin{eqnarray}
&&
{P}(a,T)=-\frac{k_BT}{\pi}\sum\limits_{l=0}^{\infty}
\left(1-\frac{1}{2}\delta_{0l}\right)\int_0^{\infty}
k_{\bot}dk_{\bot}\,q_l
\label{eq6} \\
&&
\phantom{aaa}\times
\left[\frac{r_{\|}^M(\xi_l,k_{\bot})r_{\|}^D(\xi_l,k_{\bot})}{e^{2aq_l}-
r_{\|}^M(\xi_l,k_{\bot})r_{\|}^D(\xi_l,k_{\bot})}
+\frac{r_{\bot}^M(\xi_l,k_{\bot})r_{\bot}^D(\xi_l,k_{\bot})}{e^{2aq_l}-
r_{\bot}^M(\xi_l,k_{\bot})r_{\bot}^D(\xi_l,k_{\bot})}
\right].
\nonumber
\end{eqnarray}

Using the proximity force theorem \cite{51}, one can obtain from
Eq.~(\ref{eq1}) the approximate expression for the Casimir force
acting between a sphere and a plate
\begin{equation}
F(a,T)=2\pi R {\mathcal F}(a,T).
\label{eq7}
\end{equation}
\noindent
This equation is widely used for the interpretation of measurements of the
Casimir force \cite{13,14,15,16,17,18,19,20,21,22,23,24,25,25a}.
Recently both exact analytic and numerical results for the Casimir
force in the configuration of a cylinder above a plate (electromagnetic
case) and for a sphere above a plate (scalar case) were obtained
\cite{47,48,49,50}. It was shown that the error introduced by the use of
Eq.~(\ref{eq7}) for the experimental parameters in already performed
experiments is less than 0.1\%. Using Eq.~(\ref{eq7}), the analytical 
results derived below for metal and dielectric plates, can be immediately
applied for the interpretation of measurements of the Casimir force
between Au coated sphere and Si plate \cite{25,25a,25b,25c}.

The analytic perturbation expansions in Eqs.~(\ref{eq1}) and (\ref{eq6})
can be conveniently performed by using the dimensionless variables
$\zeta_l$ and $y$
\begin{equation}
\zeta_l=\frac{\xi_l}{\omega_c}=\frac{2a\xi_l}{c}=\tau l,
\qquad
y=2aq_l,
\label{eq8}
\end{equation}
\noindent
where $\omega_c=c/(2a)$ is the characteristic frequency of the
Casimir effect and $\tau=4\pi k_BaT/(\hbar c)$.
In terms of these variables the free energy (\ref{eq1})
takes the form
\begin{eqnarray}
&&
{\mathcal F}(a,T)=\frac{\hbar c\tau}{32\pi^2a^3}\sum\limits_{l=0}^{\infty}
\left(1-\frac{1}{2}\delta_{0l}\right)\int_{\zeta_l}^{\infty}
ydy\left\{
\ln\left[1-r_{\|}^M(\zeta_l,y)r_{\|}^D(\zeta_l,y)e^{-y}
\right]\right.
\nonumber \\
&&\phantom{aaaaaaaaaaaaaaaaa}
\left.
+\ln\left[1-r_{\bot}^M(\zeta_l,y)r_{\bot}^D(\zeta_l,y)e^{-y}
\right]\right\}.
\label{eq9}
\end{eqnarray}
\noindent
Using the variables (\ref{eq8}), the reflection coefficients 
(\ref{eq2}) are
\begin{eqnarray}
&&
r_{\|}^{M,D}(\zeta_l,y)=\frac{\varepsilon_l^{M,D}y-
\sqrt{y^2+\zeta_l^2(\varepsilon_l^{M,D}-1)}}{\varepsilon_l^{M,D}y+
\sqrt{y^2+\zeta_l^2(\varepsilon_l^{M,D}-1)}},
\nonumber \\
&&
r_{\bot}^{M,D}(\zeta_l,y)=
\frac{\sqrt{y^2+\zeta_l^2(\varepsilon_l^{M,D}-1)}-
y}{\sqrt{y^2+\zeta_l^2(\varepsilon_l^{M,D}-1)}+y},
\label{eq10}
\end{eqnarray}
\noindent
where in accordance with Eq.~(\ref{eq4})
$\varepsilon_l^{M,D}=\varepsilon^{M,D}(i\zeta_l\omega_c)$.

The pressure (\ref{eq6}) is rearranged as follows:
\begin{eqnarray}
&&
{P}(a,T)=-\frac{\hbar c\tau}{32\pi^2a^4}\sum\limits_{l=0}^{\infty}
\left(1-\frac{1}{2}\delta_{0l}\right)\int_{\zeta_l}^{\infty}
y^2dy
\label{eq11} \\
&&\phantom{aaa}\times
\left[\frac{r_{\|}^M(\zeta_l,y)r_{\|}^D(\zeta_l,y)}{e^{y}-
r_{\|}^M(\zeta_l,y)r_{\|}^D(\zeta_l,y)}
+\frac{r_{\bot}^M(\zeta_l,y)r_{\bot}^D(\zeta_l,y)}{e^{y}-
r_{\bot}^M(\zeta_l,y)r_{\bot}^D(\zeta_l,y)}
\right].
\nonumber
\end{eqnarray}

The other important characteristic of the van der Waals and Casimir
interaction is the entropy
\begin{equation}
S(a,T)=-\frac{\partial{\mathcal F}(a,T)}{\partial T}.
\label{eq12}
\end{equation}
\noindent
In \cite{29,30} the behavior of the Casimir entropy at
$T\to 0$ was used as a phenomenological constraint 
on the selection of theoretically
consistent models of the dielectric response for real metals at
low frequencies. It was proposed that all consistent models
should satisfy the thermodynamic condition $S(a,0)=0$, i.e., be in agreement
with the Nernst heat theorem. In \cite{42} it was demonstrated
that this condition is respected for two dielectric plates with the
finite static dielectric permittivities. The new analytic
expressions for the free energy obtained in the present paper
permit investigate the behavior of
entropy in the configuration of one metal and one dielectric
plate and find when it vanishes with vanishing temperature.

\section{Model of ideal metal and dielectric with constant
dielectric permittivity}

To find the analytic expressions for the free energy, pressure and
entropy of the Casimir interaction between metal and dielectric,
we start from a simplified model when the metal is an ideal one and
the dielectric possesses some finite dielectric permittivity
$\varepsilon_0^D$ independent on the frequency. 
Such modeling is widely used in Casimir physics (see, e.g.,
\cite{2,3,5,6,8,10,11,12,52a}). It provides rather good description
of real metals and dielectrics at sufficiently large separations between
the interacting surfaces.
For an ideal
metal it holds $|\varepsilon^M|=\infty$ at all frequencies and
from Eq.~(\ref{eq10}) one obtains
\begin{equation}
r_{\|}^{M}(\zeta_l,y)=1,\qquad
r_{\bot}^{M}(\zeta_l,y)=1,\qquad
l\geq 0.
\label{eq13}
\end{equation}

Using Eq.~(\ref{eq13}), the free energy (\ref{eq9}) and pressure
(\ref{eq11}) are represented in a more simple form,
\begin{eqnarray}
&&
{\mathcal F}(a,T)=\frac{\hbar c\tau}{32\pi^2a^3}\sum\limits_{l=0}^{\infty}
\left(1-\frac{1}{2}\delta_{0l}\right)\int_{\zeta_l}^{\infty}
ydy\nonumber \\
&&\phantom{aaa}
\times\left\{
\ln\left[1-r_{\|}^D(\zeta_l,y)e^{-y}
\right]
+\ln\left[1-r_{\bot}^D(\zeta_l,y)e^{-y}
\right]\right\},
\nonumber \\
&&
\label{eq14}\\
&&
{P}(a,T)=-\frac{\hbar c\tau}{32\pi^2a^4}\sum\limits_{l=0}^{\infty}
\left(1-\frac{1}{2}\delta_{0l}\right)\int_{\zeta_l}^{\infty}
y^2dy
\nonumber \\
&&\phantom{aaa}\times
\left[\frac{r_{\|}^D(\zeta_l,y)}{e^{y}-r_{\|}^D(\zeta_l,y)}
+\frac{r_{\bot}^D(\zeta_l,y)}{e^{y}-r_{\bot}^D(\zeta_l,y)}
\right].
\nonumber
\end{eqnarray}
\noindent
Notice that in the framework of our model
in Eq.~(\ref{eq10}) it holds $\varepsilon_l^D=\varepsilon_0^D$,
i.e., the dielectric permittivities computed at different
imaginary Matzubara frequencies do not depend on $l$.
In particular, at $l=0$ it follows:
\begin{equation}
r_{\|}^D(0,y)=\frac{\varepsilon_0^D-1}{\varepsilon_0^D+1}
\equiv r_0, \qquad
r_{\bot}^D(0,y)=0.
\label{eq15}
\end{equation}
\noindent
In fact Eq.~(\ref{eq15}) is valid not only for our simplified model
but for any dielectric with a finite static dielectric permittivity
$\varepsilon^D(0)\equiv\varepsilon_0^D<\infty$. Usually for
nonpolar dielectrics $\varepsilon^D(i\xi)=\varepsilon_0^D=\mbox{const}$
in the frequency region from $\xi=0$ up to rather high frequencies of about
$10^{15}\,$rad/s and for higher frequencies 
$\varepsilon^D(i\xi)$ decreases to unity.
The simplified model does not take the latter into account (in the 
next section we show that this does not influence the first terms in
the asymptotic behavior of the free energy, entropy and pressure
at low temperature).

Now we proceed with the derivation of the asymptotic behavior of
the Casimir free energy at low temperature ($\tau\ll 1$).
Using the Abel-Plana formula \cite{3,8}
\begin{equation}
\sum\limits_{l=0}^{\infty}\left(1-\frac{1}{2}\delta_{l0}\right)
F(l)=\int_0^{\infty}F(t)dt+
i\int_0^{\infty}dt\frac{F(it)-F(-it)}{e^{2\pi t}-1},
\label{eq16}
\end{equation}
\noindent
where $F(z)$ is an analytic function in the right-plane,
we can rearrange Eq.~(\ref{eq14}) to the form
\begin{equation}
{\mathcal F}(a,T)=E(a)+\Delta{\mathcal F}(a,T).
\label{eq17}
\end{equation}
\noindent
Here,
\begin{equation}
E(a)=\frac{\hbar c}{32\pi^2a^3}\int_{0}^{\infty}d\zeta
\int_{\zeta}^{\infty}f(\zeta,y)dy
\label{eq18}
\end{equation}
\noindent
is the energy of the Casimir interaction at zero temperature,
\begin{equation}
\Delta{\mathcal F}(a,T)=\frac{i\hbar c\tau}{32\pi^2a^3}
\int_{0}^{\infty}dt
\frac{F(i\tau t)-F(-i\tau t)}{e^{2\pi t}-1}
\label{eq19}
\end{equation}
\noindent
is the thermal correction to it, and the following notations are
introduced,
\begin{eqnarray}
&&
f(\zeta,y)=y\ln\left[1-r_{\|}^D(\zeta,y)e^{-y}\right]+
y\ln\left[1-r_{\bot}^D(\zeta,y)e^{-y}\right],
\nonumber \\
&&
F(x)=\int_{x}^{\infty}dyf(x,y).
\label{eq20}
\end{eqnarray}

The expansion of $f(x,y)$ in powers of $x$ takes the form
\begin{eqnarray}
&&
f(x,y)=y\ln(1-r_0e^{-y})
\label{eq21} \\
&&\phantom{aaaaaaa}
-x^2\left(\frac{\varepsilon_0^D-1}{4y}e^{-y}-
\frac{\varepsilon_0^D}{\varepsilon_0^D+1}\sum\limits_{n=1}^{\infty}
r_0^n\frac{e^{-ny}}{y}\right)+\mbox{O}(x^3).
\nonumber
\end{eqnarray}
\noindent
To find $F(x)$ in Eq.~(\ref{eq20}) we integrate the right-hand side of
Eq.~(\ref{eq21}) with respect to $y$. Notice that the first term
on the right-hand side of Eq.~(\ref{eq21}) does not contribute to the
first expansion orders of $F(ix)-F(-ix)$ which is in fact the
quantity of our interest.
This is because in the expression
\begin{equation}
\int_{x}^{\infty}ydy\ln(1-r_0e^{-y})=\int_{0}^{\infty}vdv
\ln(1-r_0e^{-v})+\mbox{O}(x^2),
\label{eq22}
\end{equation}
\noindent
where the new variable $v=y-x$ was introduced, the first-order in
$x$ contribution vanishes. Thus, this term could contribute to
$F(ix)-F(-ix)$ only starting from the third expansion order.
Integrating the second term on the right-hand side of Eq.~(\ref{eq21})
using the formula
\begin{equation}
\int_{x}^{\infty}dy
\frac{e^{-ny}}{y}=-\mbox{Ei}(-nx),
\label{eq23}
\end{equation}
\noindent
where Ei$(z)$ is the exponential integral function, we finally
obtain
\begin{equation}
F(ix)-F(-ix)=i\pi\frac{(\varepsilon_0^D-1)^2}{4(\varepsilon_0^D+1)}x^2
-i\gamma x^3+\mbox{O}(x^4),
\label{eq24}
\end{equation}
\noindent
where the third order real expansion coefficient $\gamma$ cannot be
determined at this stage of our calculations because all powers
in the expansion of $f(x,y)$ in powers of $x$ contribute to it.

Now we substitude Eq.~(\ref{eq24}) in Eq.~(\ref{eq19}) and find
the free energy (\ref{eq17})
\begin{equation}
{\mathcal F}(a,T)=E(a)-\frac{\hbar c}{32\pi^2a^3}\left[
\frac{\zeta(3)}{16\pi^2}
\frac{(\varepsilon_0^D-1)^2}{\varepsilon_0^D+1}\tau^3-K_4\tau^4+
\mbox{O}(\tau^5)\right],
\label{eq25}
\end{equation}
\noindent
where $K_4=\gamma/240$ and $\zeta(z)$ is the Riemann zeta function.

The Casimir pressure is obtained from Eqs.~(\ref{eq5}) and (\ref{eq25}).
It is equal to
\begin{equation}
P(a,T)=P_0(a)-\frac{\hbar c}{32\pi^2a^4}\left[
K_4\tau^4+\mbox{O}(\tau^5)\right],
\label{eq26}
\end{equation}
\noindent
where $P_0(a)=-\partial E(a)/\partial a$.

In order to determine the coefficients $K_4$, we now start
from the Lifshitz representation of the pressure in Eq.~(\ref{eq14}).
Using the Abel-Plana formula (\ref{eq16}), we rearrange
Eq.~(\ref{eq14}) to the form analogical to (\ref{eq17})--(\ref{eq19}),
\begin{eqnarray}
&&
P(a,T)=P_0(a)+\Delta P(a,T),
\nonumber \\  
&&
P_0(a)=\frac{3\hbar c}{32\pi^2a^4}\int_{0}^{\infty}d\zeta
\int_{\zeta}^{\infty}f(\zeta,y)dy,
\label{eq27} \\
&&
\Delta P(a,T)=-\frac{i\hbar c\tau}{32\pi^2a^4}
\int_{0}^{\infty}dt\frac{\Phi(i\tau t)-\Phi(-i\tau t)}{e^{2\pi t}-1}.
\nonumber
\end{eqnarray}
\noindent
Here $P_0(a)$ is the Casimir pressure at zero temperature,
$\Delta P(a,T)$ is the thermal correction to it and the following
notation is introduced:
\begin{equation}
\Phi_{\|,\bot}(x)=\int_{x}^{\infty}dy
\frac{y^2r_{\|,\bot}(x,y)}{e^{y}-r_{\|,\bot}(x,y)}.
\label{eq28}
\end{equation}

To find the expansion of $\Phi(ix)-\Phi(-ix)$ in powers of $x$,
we first deal with $\Phi_{\bot}(x)$. By adding and subtracting the
asymptotic behavior of the intergrand function at small $x$,
\begin{equation}
\frac{y^2r_{\bot}(x,y)}{e^{y}-r_{\bot}(x,y)}=
\frac{1}{4}(\varepsilon_0^D-1)x^2e^{-y}+\mbox{O}(x^3),
\label{eq29}
\end{equation}
under the integral in Eq.~(\ref{eq28}) and introducing the new
variable $v=y/x$, the function $\Phi_{\bot}(x)$ can be identically
rearranged and expanded in powers of $x$ as follows:
\begin{eqnarray}
&&
\Phi_{\bot}(x)=\frac{1}{4}(\varepsilon_0^D-1)x^2e^{-x}+x^3
\int_{1}^{\infty}dv\left[v^2\sum\limits_{n=1}^{\infty}
r_{\bot}^{n}(v)e^{-nvx}-\frac{1}{4}(\varepsilon_0^D-1)e^{-vx}\right]
\nonumber \\
&&
\phantom{\Phi_{\bot}(x)}
=\frac{1}{4}(\varepsilon_0^D-1)x^2(1-x)
\label{eq30}\\
&&\phantom{aa}
+x^3
\int_{1}^{\infty}dv\left[\frac{v^2r_{\bot}(v)}{1-r_{\bot}(v)}-
\frac{\varepsilon_0^D-1}{4}\right]
+\mbox{O}(x^4).
\nonumber
\end{eqnarray}
\noindent
Calculating the integral on the right-hand side of Eq.~(\ref{eq30}),
we arrive at the result
\begin{equation}
\Phi_{\bot}(x)=\frac{\varepsilon_0^D-1}{4}x^2-\frac{1}{6}
\left(\varepsilon_0^D\sqrt{\varepsilon_0^D}-1\right)x^3
+\mbox{O}(x^4).
\label{eq31}
\end{equation}

To deal with $\Phi_{\|}(x)$, we add and subtract under the integral in
Eq.~(\ref{eq28}) the two first expansion terms of the integrated function
in powers of $x$,
\begin{eqnarray}
&&
\Phi_{\|}(x)=\int_{x}^{\infty}y^2dy\left[\frac{r_0}{e^y-r_0}-
\frac{\varepsilon_0^Dr_0e^{-y}x^2}{y^2(\varepsilon_0^D+
1)(1-r_0e^{-y})^2}\right]
\nonumber \\
&&
\phantom{\Phi_{\|}(x)}
+\int_{x}^{\infty}y^2dy\left[\frac{r_{\|}(x,y)}{e^y-r_{\|}(x,y)}
-\frac{r_0}{e^y-r_0}+
\frac{\varepsilon_0^Dr_0e^{-y}x^2}{y^2(\varepsilon_0^D+
1)(1-r_0e^{-y})^2}\right].
\label{eq32}
\end{eqnarray}
\noindent
The asymptotic expansions of the first and second integrals on
the right-hand side of Eq.~(\ref{eq32}) are
\begin{eqnarray}
&&
2\mbox{Li}_3(r_0)-
\frac{\varepsilon_0^D(\varepsilon_0^D-1)}{2(\varepsilon_0^D+1)}x^2
+\frac{1}{12}(\varepsilon_0^D-1)(3\varepsilon_0^D-2)x^3+
\mbox{O}(x^4),
\label{eq33} \\
&&
\left[-\frac{1}{4}\varepsilon_0^D(\varepsilon_0^D-1)-
\frac{1}{6}\varepsilon_0^D(\varepsilon_0^D\sqrt{\varepsilon_0^D}-1)+
\frac{1}{2}\varepsilon_0^D(\varepsilon_0^D-1)\sqrt{\varepsilon_0^D}
\right]x^3
\nonumber \\
&&\phantom{aaaaaaaaaaaaaaaaaaaaaaaaaaaaaaa}
+\mbox{O}(x^4),
\label{eq34}
\end{eqnarray}
\noindent
respectively, where Li${}_n(z)$ is the polylogarithm function.
By summing Eqs.~(\ref{eq33}) and (\ref{eq34}) we find
\begin{eqnarray}
&&
\Phi_{\|}(x)=2\mbox{Li}_3(r_0)-
\frac{\varepsilon_0^D(\varepsilon_0^D-1)}{2(\varepsilon_0^D+1)}x^2
\label{eq35} \\
&&
\phantom{\Phi_{\|}(x)}-\frac{1}{6}\left[(\varepsilon_0^D-1)+
(\varepsilon_0^D\sqrt{\varepsilon_0^D}-1)-
3\varepsilon_0^D(\varepsilon_0^D-1)\sqrt{\varepsilon_0^D}
\right]x^3+
\mbox{O}(x^4).
\nonumber
\end{eqnarray}
\noindent
After summing Eqs.~(\ref{eq31}) and (\ref{eq35}) the following result
is obtained:
\begin{equation}
\Phi(ix)-\Phi(-ix)=-\frac{2i}{3}\left(1-
2\varepsilon_0^D\sqrt{\varepsilon_0^D}+
\left(\varepsilon_0^D\right)^2
\sqrt{\varepsilon_0^D}\right)x^3+
\mbox{O}(x^4).
\label{eq36}
\end{equation}

Substituting this in Eq.~(\ref{eq27}) and integrating we arrive at the 
asymptotic expression
for the Casimir pressure in the limit of small $\tau$,
\begin{equation}
P(a,T)=P_0(a)-\frac{\hbar c}{32\pi^2a^4}\left[
\frac{1-2\varepsilon_0^D\sqrt{\varepsilon_0^D}+
\left(\varepsilon_0^D\right)^2\sqrt{\varepsilon_0^D}}{360}
\tau^4+\mbox{O}(\tau^5)\right].
\label{eq37}
\end{equation}
\noindent
The comparison of this equation with Eq.~(\ref{eq26}) leads to the 
explicit expression
for the coefficient $K_4$:
\begin{equation}
K_4=\frac{1}{360}\left(1-2\varepsilon_0^D\sqrt{\varepsilon_0^D}+
\left(\varepsilon_0^D\right)^2\sqrt{\varepsilon_0^D}\right)
\label{eq38}
\end{equation}
\noindent
and, thus, the explicit asymptotic expression (\ref{eq25}) for the free 
energy is also
fully determined.

Notice that the energy and pressure at zero temperature [$E(a)$ and $P_0(a)$ 
defined
in Eqs.~(\ref{eq18}) and (\ref{eq27}), respectively] depend on the separation 
through
only the factors $a^{-3}$ and $a^{-4}$ in front of the integrals. They can be
conveniently presented in the form
\begin{eqnarray}
&&
E(a)=-\frac{\pi^2}{720}\,\frac{\hbar c}{a^3}\,\psi_{DM}(\varepsilon_0^D),
\label{eq39} \\
&&
P_0(a)=-\frac{\pi^2}{240}\,\frac{\hbar c}{a^4}\,\psi_{DM}(\varepsilon_0^D),
\nonumber
\end{eqnarray}
\noindent
where the function $\psi_{DM}(\varepsilon_0)$ is defined as
\begin{equation}
\psi_{DM}(\varepsilon_0^D)=-\frac{45}{2\pi^4}\int_{0}^{\infty}d\zeta
\int_{\zeta}^{\infty}f(\zeta,y)dy.
\label{eq40}
\end{equation}
\noindent
In fact, $\psi_{DM}$ in Eqs.~(\ref{eq39}), (\ref{eq40}) is the correction 
factor
to the famous Casimir result \cite{1} obtained for two ideal metals.
It is equal to the function $\varphi_{DM}$ introduced in \cite{12},
multipled by $r_0$.

The function $\psi_{DM}(\varepsilon_0^D)$ in Eq.~(\ref{eq40}) can be 
presented in a more simple analytical form as follows. 
Presenting the logarithms in Eq.~(\ref{eq20}) 
as series and
changing the order of integrations, one obtains
\begin{eqnarray}
&&
\psi_{DM}(\varepsilon_0^D)=\frac{45}{2\pi^4}\sum\limits_{n=1}^{\infty}
\frac{1}{n}\int_{0}^{\infty}ydye^{-ny}
\label{eq40a} \\
&&
\phantom{\psi_{DM}(\varepsilon_0^D)}\times
\int_{0}^{y}d\zeta\left\{\left[r_{\|}^D(\zeta,y)\right]^n+
\left[r_{\bot}^D(\zeta,y)\right]^n\right\}.
\nonumber
\end{eqnarray}
\noindent
Introducing the new variable $w=\zeta/y$, we rearrange Eq.~(\ref{eq40a})
to the form
\begin{eqnarray}
&&
\psi_{DM}(\varepsilon_0^D)=\frac{45}{2\pi^4}\sum\limits_{n=1}^{\infty}
\frac{1}{n}\int_{0}^{\infty}y^2dye^{-ny}
\label{eq40b} \\
&&
\phantom{\psi_{DM}(\varepsilon_0^D)}\times
\int_{0}^{1}dw\left\{\left[r_{\|}^D(w)\right]^n+
\left[r_{\bot}^D(w)\right]^n\right\},
\nonumber
\end{eqnarray}
\noindent
where
\begin{eqnarray}
&&
r_{\|}^D(w)=\frac{\varepsilon_0^D-
\sqrt{1+(\varepsilon_0^D-1)w^2}}{\varepsilon_0^D+
\sqrt{1+(\varepsilon_0^D-1)w^2}},
\nonumber \\
&&
r_{\bot}^D(w)=\frac{\sqrt{1+(\varepsilon_0^D-1)w^2}-
1}{\sqrt{1+(\varepsilon_0^D-1)w^2}+1}.
\label{eq40c}
\end{eqnarray}
\noindent
Calculating the integral in $y$ and performing the summation with
respect  to $n$ in Eq.~(\ref{eq40b}) one arrives at
\begin{equation}
\psi_{DM}(\varepsilon_0^D)=\frac{45}{\pi^4}\int_{0}^{\infty}
dw\left\{\mbox{Li}_4\left[r_{\|}^D(w)\right]+
\mbox{Li}_4\left[r_{\bot}^D(w)\right]\right\}.
\label{eq40d}
\end{equation}
\noindent
In Fig.~1 the function (\ref{eq40d}) is plotted versus
$\varepsilon_0^D$ as a solid line (when $\varepsilon_0^D\to 1$
it goes to zero and when $\varepsilon_0^D\to\infty$ it
goes to unity reproducing the limit of ideal metals).

It is notable that the model under consideration represents
correctly the Casimir energy and pressure (\ref{eq39}) at $T=0$ in only
the retarded regime (i.e., at sufficiently large separations).
As to the thermal corrections in Eqs.~(\ref{eq25}) and (\ref{eq37}),
the obtained expressions are valid also at short separations in a
nonretarded regime under the condition that the parameter $\tau$ is
sufficiently small due to sufficiently low temperature.

From Eqs.~(\ref{eq12}) and (\ref{eq25}) the asymptotic behavior of the
Casimir entropy in the limit of small $\tau$ is given by
\begin{eqnarray}
&&
S(a,T)=\frac{3k_B\zeta(3)\left(\varepsilon_0^D-
1\right)^2}{128\pi^3a^2\left(\varepsilon_0^D+1\right)}\tau^2
\label{eq41} \\
&&
\phantom{S(a,T)}\times
\left[1-\frac{8\pi^2\left(\varepsilon_0^D+1\right)\left(1-
2\varepsilon_0^D\sqrt{\varepsilon_0^D}+
\left(\varepsilon_0^D\right)^2
\sqrt{\varepsilon_0^D}\right)}{135\zeta(3)\left(\varepsilon_0^D-
1\right)^2}\tau+\mbox{O}(\tau^2)\right].
\nonumber
\end{eqnarray}
\noindent
As is seen from Eq.~(\ref{eq41}), the entropy of the Casimir
interaction between metal and dielectric plates vanishes with
vanishing temperature as is required by the Nernst heat theorem
(note that the first term of order $\tau^2$ in Eq.~(\ref{eq41})
was obtained in \cite{46}). The important property of the
perturbation expansions in powers of $\tau$ in Eqs.~(\ref{eq25}),
(\ref{eq37}) and (\ref{eq41}) is that it is impermissible
to consider the limiting case $\varepsilon_0^D\to\infty$ in
order to obtain the case of two ideal metals like it was
discussed above in application to Eq.~(\ref{eq39}). The mathematical
reason is that in the power expansion of functions depending on
$\varepsilon_0^D$ as a parameter the limiting transitions
$\varepsilon_0^D\to\infty$ and $\tau\to 0$ are not interchangeable.
Of great importance is the possibility to apply Eq.~(\ref{eq41}) at
as small $T$ as is wished. This is the principal advantage of analytical 
calculations as compared to numerical ones.

Now we consider the opposite limiting case $\tau\gg 1$, i.e., the
limit of high temperatures (large separations). Here the main
contribution to the free energy (\ref{eq14}) is given by the term
with $l=0$ whereas all terms with $l\geq 1$ are
exponentially small \cite{8},
\begin{equation}
{\mathcal F}(a,T)=\frac{\hbar c\tau}{64\pi^2a^3}
\int_{0}^{\infty}ydy\ln\left(1-r_0e^{-y}\right).
\label{eq42}
\end{equation}
\noindent
By integrating in Eq.~(\ref{eq42}) we obtain
\begin{equation}
{\mathcal F}(a,T)=-\frac{k_BT}{16\pi a^2}
\mbox{Li}_3(r_0).
\label{eq43}
\end{equation}
\noindent
For the Casimir pressure and entropy at $\tau\gg 1$ from
Eqs.~(\ref{eq5}), (\ref{eq12}) and (\ref{eq43}) it follows
\begin{equation}
P(a,T)=-\frac{k_BT}{8\pi a^3}
\mbox{Li}_3(r_0), \qquad
S(a,T)=\frac{k_B}{16\pi a^2}
\mbox{Li}_3(r_0).
\label{eq44}
\end{equation}

\section{Thermal Casimir force between metal and dielectric
with {\protect \\} frequency-dependent dielectric permittivities}

In this section we obtain the analytic expressions for the 
low-temperature behavior of the Casimir interaction between metal
and dielectric plates taking into account the dependence of their
dielectric permittivities on the frequency. The metal plate is
described by the dielectric permittivity of the plasma model,
\begin{equation}
\varepsilon^M(i\xi_l)=1+\frac{\omega_p^2}{\xi_l^2},
\label{eq45}
\end{equation}
\noindent
where $\omega_p=2\pi c/\lambda_p$ is the plasma frequency, and
$\lambda_p$ is the plasma wavelength. In the theory of the thermal
Casimir force this description was first used in \cite{34,35}
and was shown to work good at separations between plates greater 
than the plasma wavelength. At such separations the characteristic 
frequency of the Casimir effect $\omega_c$ belongs to the region of
infrared optics where the relaxation processes do not play any
role \cite{52b}.

For dielectric plate we use the Ninham-Parsegian representation of the
dielectric permittivity along the imaginary frequency axis \cite{52,53},
\begin{equation}
\varepsilon^D(i\xi_l)=1+\sum\limits_{j}
\frac{C_j}{1+\frac{\xi_l^2}{\omega_j^2}}.
\label{eq46}
\end{equation}
\noindent
Here $C_j$ are the absorption strengths satisfying the condition
\begin{equation}
\sum\limits_{j}C_j=\varepsilon_0^D-1,
\label{eq47}
\end{equation}
\noindent
and $\omega_j$ are the characteristic absorption frequencies 
[recall that now $\varepsilon_0^D=\varepsilon^D(0)<\infty$].
Eq.~(\ref{eq46}) gives a very accurate approximate description
of the dielectric properties for many dielectrics. It has been
successfully used by many authors for the comparison of
experimental data with theory \cite{54}.

{}From Eq.~(\ref{eq46}) we return to the same values (\ref{eq15})
of the reflection coefficients of the dielectric plate
at zero frequency as were obtained
in the simplified model of the frequency-independent dielectric
permittivity. Thus, due to the zero value of $r_{\bot}^D(0,y)$,
the transverse electric mode at zero
frequency does not contribute to the free energy (\ref{eq9})
of the Casimir interaction between metal and dielectric regardless
of the value of $r_{\bot}^M(0,y)$ for a metal. 
As was told in the Introduction, there are different approaches on
how to correctly calculate the transverse electric coefficient
at zero frequency, $r_{\bot}^M(0,y)$, for a 
plate made of real metal. In the configuration of metal and
dielectric this problem, however, does not influence the result.
Note that if we would use instead of Eq.~(\ref{eq45}) the Drude model,
taking relaxation into account, the prime perturbation orders in all
results below remain unchanged for metals with perfect crystal
lattices. The role of impurities in the validity of the Nernst heat
theorem in the case of two metal plates is discussed in 
\cite{28,30,38,39,40,41,55a}.

We start from Eq.~(\ref{eq9}) for the free energy. Once again,
using the Abel-Plana
formula, Eq.~(\ref{eq9}) can be represented by Eq.~(\ref{eq17})
as the sum of $\hat{E}(a)$ in Eq.~(\ref{eq18}) and
$\Delta\hat{\mathcal F}(a,T)$ in Eqs.~(\ref{eq19}) and Eq.~(\ref{eq20}),
where we mark by a hat all quantities related to {\it real} metal and 
dielectric.
The single difference is that the function $f(x,y)$ in Eq.~(\ref{eq20})
should be replaced by
\begin{eqnarray}
&&
\hat{f}(x,y)\equiv\hat{f}_{\|}(x,y)+\hat{f}_{\bot}(x,y),
\label{eq48} \\
&&
\hat{f}_{\|,\bot}(x,y)=y\ln\left[1-\hat{r}_{\|,\bot}^M(x,y)
\hat{r}_{\|,\bot}^D(x,y)e^{-y}\right].
\nonumber
\end{eqnarray}
\noindent
It is notable that for real metal and dielectric 
$\hat{E}(a)$  and
$\Delta\hat{\mathcal F}(a,T)$ in Eq.~(\ref{eq17})
may lose the obvious meaning of the energy at zero temperature
and the thermal correction to it. In fact, this meaning is preserved
only in the case when the dielectric permittivities
$\varepsilon^{M,D}(i\xi)$ do not depend on the temperature as 
a parameter like it was in Section~3. In the latter case it holds
\begin{equation}
\Delta{\mathcal F}(a,T)={\mathcal F}(a,T)-{\mathcal F}(a,0)={\mathcal F}(a,T)-E(a)
\label{eq49}
\end{equation}
\noindent
in accordance with the intuitive definition of the thermal correction.
If, however, $\varepsilon^{M}(i\xi)$ or $\varepsilon^{D}(i\xi)$
or both depend on the temperature as a parameter, Eq.~(\ref{eq49}) is 
violated. In this case $\Delta\hat{\mathcal F}(a,T)$ defined in 
Eq.~(\ref{eq19}) takes into account only the part of temperature
dependence of the free energy originating from the Matsubara
frequencies and is not equal to 
$\hat{\mathcal F}(a,T)-\hat{\mathcal F}(a,0)$. Moreover, in this case
$\hat{E}(a)$ in Eqs.~(\ref{eq17}) and (\ref{eq18}) is in fact
temperature-dependent and it would be more correct to use the
notation $\hat{E}=\hat{E}(a,T)$.

To obtain the analytic expressions of our interest we develop the
perturbation theory in two small parameters $\tau$ and
$\eta\equiv\delta/(2a)$, where $\delta=\lambda_p/(2\pi)$ is
the penetration depth of the electromagnetic oscillations into
a metal. For the sake of simplicity we will consider dielectrics
which can be described by Eq.~(\ref{eq46}) with only one
oscillator, i.e., with $j=1$. The high-resistivity Si is a typical
example of such materials. The function $F(x)$ in Eq.~(\ref{eq20})
can be conveniently presented in the form
\begin{equation}
\hat{F}(x)=\hat{F}_{\|}(x)+\hat{F}_{\bot}(x),
\qquad
\hat{F}_{\|,\bot}(x)=\int_{x}^{\infty}dy\hat{f}_{\|,\bot}(x,y).
\label{eq50}    
\end{equation}
\noindent
As a first step we perform the expansion with respect to the powers
of small parameter $\eta$. This results in:
\begin{eqnarray}
&&
\hat{F}_{\|}(x)=\int_{x}^{\infty}ydy
\ln\left[1-\hat{r}_{\|}^D(x,y)e^{-y}\right]
+2x^2\eta\int_{x}^{\infty}dy
\frac{\hat{r}_{\|}^D(x,y)}{e^y-\hat{r}_{\|}^D(x,y)}
\nonumber \\
&&\phantom{\hat{F}_{\|}(x)}
-2x^4\eta^2\int_{x}^{\infty}dy
\frac{e^{y}\hat{r}_{\|}^D(x,y)}{y[e^y-\hat{r}_{\|}^D(x,y)
]^2}+\mbox{O}(\eta^3),
\nonumber \\
&&\label{eq51} \\
&&
\hat{F}_{\bot}(x)=\int_{x}^{\infty}ydy
\ln\left[1-\hat{r}_{\bot}^D(x,y)e^{-y}\right]
+2\eta\int_{x}^{\infty}y^2dy
\frac{\hat{r}_{\bot}^D(x,y)}{e^y-\hat{r}_{\bot}^D(x,y)}
\nonumber \\
&&\phantom{\hat{F}_{\|}(x)}
-2\eta^2\int_{x}^{\infty}y^3dy
\frac{e^{y}\hat{r}_{\bot}^D(x,y)}{[e^y-\hat{r}_{\bot}^D(x,y)
]^2}+\mbox{O}(\eta^3).
\nonumber
\end{eqnarray}
\noindent
The dielectric reflection coefficients in Eq.~(\ref{eq51}) are obtained 
after the substitution of Eqs.~(\ref{eq46}) and (\ref{eq47}) with $j=1$ 
in Eq.~(\ref{eq10}): 
\begin{eqnarray}
&&
\hat{r}_{\|}^D(x,y)=
\frac{\left(1+\frac{\varepsilon_0^D-1}{1+b^2x^2}\right)y-\sqrt{y^2+
\frac{\varepsilon_0^D-1}{1+b^2x^2}x^2}}{\left(1+
\frac{\varepsilon_0^D-1}{1+b^2x^2}\right)y+\sqrt{y^2+
\frac{\varepsilon_0^D-1}{1+b^2x^2}x^2}},
\nonumber \\
&&
\hat{r}_{\bot}^D(x,y)=
\frac{\sqrt{y^2+\frac{\varepsilon_0^D-1}{1+b^2x^2}x^2}-
y}{\sqrt{y^2+\frac{\varepsilon_0^D-1}{1+b^2x^2}x^2}+y},
\label{eq52}
\end{eqnarray}
\noindent
where $b\equiv\omega_c/\omega_1$.

Let us consecutively consider the contributions to $\hat{F}(x)$ from
the terms of order $\eta^0$, $\eta$ and $\eta^2$ in Eq.~(\ref{eq51}).
As to the terms of order $\eta^0$ [the first and fourth lines
in Eq.~(\ref{eq51})], the expansion in powers of small $x$ (small
$\tau$) performed using Eq.~(\ref{eq52}) leads to
\begin{eqnarray}
&&
\hat{F}_{\eta^0}(x)=F(x)-b^2x^4\left\{
\frac{3\left(\varepsilon_0^D\right)^2+
2\varepsilon_0^D-1}{(\varepsilon_0^D+1)^2}
\sum\limits_{n=1}^{\infty}nr_0^n\mbox{Ei}(-nx)\right.
\label{eq53} \\
&&
\phantom{\hat{F}_{\eta^0}(x,y)}\left.
-r_0^2\sum\limits_{n=1}^{\infty}nr_0^n\mbox{Ei}\left[-(n+1)x\right]
+\frac{\varepsilon_0^D-1}{4}\mbox{Ei}(-x)\right\}
+\mbox{O}(x^5).
\nonumber
\end{eqnarray}
\noindent
Here $F(x)$ was already calculated in Sec.~III and results in
Eq.~(\ref{eq24}). The additional contributions to the right-hand
side of Eq.~(\ref{eq53}) lead to the term of order $\tau^4$
in $\hat{F}_{\eta^0}(i\tau t)-\hat{F}_{\eta^0}(-i\tau t)$ 
and of order $\tau^5$
in the free energy. Thus, they can be omitted (like in Sec.~III, 
we preserve only the terms of order $\tau^3$ and $\tau^4$).

To find the contribution to $\hat{F}(x)$ of order $\eta$ [we 
use the notation 
$\hat{F}_{\eta}(x)$], we expand in powers of $x$ the following
quantities under the integrals in Eq.~(\ref{eq51}):
\begin{eqnarray}
&&
\frac{x^2\hat{r}_{\|}^D(x,y)}{e^y-\hat{r}_{\|}^D(x,y)}
=x^2\frac{r_0}{e^y-r_0}-x^4
\frac{e^yr_0\varepsilon_0^D}{y^2(\varepsilon_0^D+1)(e^y-r_0)^2}
\nonumber \\
&&\phantom{aaaaaaaaaaaa}
-2b^2x^4\frac{r_0e^y}{(\varepsilon_0^D+1)(e^y-r_0)^2}
+\mbox{O}(x^5),
\label{eq54} \\
&&
\frac{y^2\hat{r}_{\bot}^D(x,y)}{e^y-\hat{r}_{\bot}^D(x,y)}
=x^2\frac{(\varepsilon_0^D-1)e^{-y}}{4}-x^4
\frac{(\varepsilon_0^D-1)^2e^{-2y}(2e^y-1)}{16y^2}
\nonumber \\
&&\phantom{aaaaaaaaaaaa}
-b^2x^4\frac{(\varepsilon_0^D-1)e^{-y}}{4}
+\mbox{O}(x^5).
\nonumber
\end{eqnarray}
\noindent
By integrating of the third terms on the right-hand side of
equations (\ref{eq54}) with respect to $y$ from $x$ to infinity,
we find that they contribute to 
$\hat{F}_{\eta}(i\tau t)-\hat{F}_{\eta}(-i\tau t)$ 
only in the order $\tau^5$ and, thus, to the free energy
in the order $\tau^6$. Because of this they can be omitted.
The integration of the first two terms on the right-hand side of
equations (\ref{eq54}) leads to
\begin{eqnarray}
&&
\hat{F}_{\eta,\|}(i\tau t)-\hat{F}_{\eta,\|}(-i\tau t)=
i\eta\tau^3t^3(\varepsilon_0^D-1)(\varepsilon_0^D+2),
\label{eq55} \\
&&
\hat{F}_{\eta,\bot}(i\tau t)-\hat{F}_{\eta,\bot}(-i\tau t)=
i\eta\tau^3t^3
\frac{(\varepsilon_0^D-1)(\varepsilon_0^D+3)}{4}.
\nonumber
\end{eqnarray}
\noindent
From Eq.~(\ref{eq55}) it follows
\begin{equation} 
\hat{F}_{\eta}(i\tau t)-\hat{F}_{\eta}(-i\tau t)=
i\eta\tau^3t^3
\frac{(\varepsilon_0^D-1)(5\varepsilon_0^D+11)}{4}.
\label{eq56}
\end{equation}
\noindent
As to the terms of order $\eta^2$ in Eq.~(\ref{eq51}), their lowest
order contributions to 
$\hat{F}_{\eta^2,\|}(i\tau t)-\hat{F}_{\eta^2,\|}(-i\tau t)$ and to
$\hat{F}_{\eta^2,\bot}(i\tau t)-\hat{F}_{\eta^2,\bot}(-i\tau t)$
are of order $\tau^4$ and $\tau^5$, respectively. This leads to
the respective contributions of order $\tau^5$ and $\tau^6$ to
the free energy which we omit in our analysis.

Using Eq.~(\ref{eq19}), the respective correction to the Casimir
free energy takes the form
\begin{equation}
\Delta\hat{\mathcal F}_{\eta}(a,T)=-\frac{\hbar c}{30720\pi^2a^3}
\eta\tau^4(\varepsilon_0^D-1)(5\varepsilon_0^D+11).
\label{eq57}
\end{equation}
\noindent
Remarkably, $\eta\tau^4\sim a^3$ and the correction (\ref{eq57})
does not depend on the separation. Thus, there is no correction
to the Casimir pressure of order $\eta\tau^q$ with $q\leq 4$
due to the finite conductivity of a metal plate. Recall that
in the configuration of two ideal metal plates the main thermal
correction at low temperature is of order $\tau^4$.
If the nonideality of a metal is taken into account, the
correction of order $\tau^3$ arises \cite{8}. From this it follows
that the thermal correction in the configuration metal-dielectric
is less sensitive to the finite conductivity of a metal than in
configuration of two metals.

Combining the contributions from the zeroth and first orders in
$\eta$ in Eqs.~(\ref{eq25}) and (\ref{eq57}), the free energy
at low temperatures for the configuration or real metal and
real dielectric is
\begin{eqnarray}
&&
\hat{\mathcal F}(a,T)=\hat{E}(a)-\frac{\hbar c}{32\pi^2a^3}\left[
\frac{\zeta(3)}{16\pi^2}
\frac{(\varepsilon_0^D-1)^2}{\varepsilon_0^D+1}\tau^3-K_4\tau^4
\right.
\label{eq58} \\
&&
\phantom{\hat{\mathcal F}(a,T)}\left.
\vphantom{\frac{\zeta(3)}{6\pi^2}\frac{(\varepsilon_0^D)^2}{\varepsilon_0^D}}
+\frac{1}{960}(\varepsilon_0^D-1)(5\varepsilon_0^D+11)\eta\tau^4
+\mbox{O}(\tau^5)\right],
\nonumber
\end{eqnarray}
\noindent
where $K_4$ is defined in Eq.~(\ref{eq38}). It is notable that the
low-temperature behavior of the free energy is not influenced by
the absorption bands of the dielectric material and are determined
by only the static dielectric permittivity. This is in analogy to
the case of two dielectric plates \cite{42,43}. For the Casimir
pressure between plates made of real metal and dielectric
Eq.~(\ref{eq37}) is preserved with the replacement of $P_0(a)$
for $\hat{P}_0(a)$ given below.

Now we derive the analytic representation for the Casimir energy
$\hat{E}(a)$ in the configuration with one plate made of real metal
and another plate made of real dielectric. Expanding in powers of 
$\eta$ in Eq.~(\ref{eq18}), with $f$ replaced by $\hat{f}$ from
Eq.~(\ref{eq48}), we obtain  
\begin{eqnarray}
&&
\hat{E}(a)=\frac{\hbar c}{32\pi^2a^3}\left\{
\vphantom{\int_{\zeta}^{\infty}y^3dy\frac{e^{y}
\hat{r}_{\bot}^D(\zeta,y)}{\left(e^y-
\hat{r}_{\bot}^D(\zeta,y)\right)^2}}
\int_{0}^{\infty}d\zeta
\int_{\zeta}^{\infty}ydy\left[
\ln\left(1-\hat{r}_{\|}^D(\zeta,y)e^{-y}\right)\right.\right.
\nonumber \\
&&
\phantom{\hat{E}(a)}
\left.+
\ln\left(1-\hat{r}_{\bot}^D(\zeta,y)e^{-y}\right)\right]
\label{eq59} \\
&&
\phantom{\hat{E}(a)}
+2\eta\int_{0}^{\infty}d\zeta\biggl[
\zeta^2\int_{\zeta}^{\infty}dy
\frac{\hat{r}_{\|}^D(\zeta,y)}{e^y-\hat{r}_{\|}^D(\zeta,y)}+
\int_{\zeta}^{\infty}y^2dy
\frac{\hat{r}_{\bot}^D(\zeta,y)}{e^y-\hat{r}_{\bot}^D(\zeta,y)}
\biggr]
\nonumber \\
&&
\phantom{\hat{E}(a)}
\left.
-2\eta^2\int_{0}^{\infty}d\zeta\biggl[
\zeta^4\int_{\zeta}^{\infty}dy
\frac{e^{y}\hat{r}_{\|}^D(\zeta,y)}{y\bigl(e^y-
\hat{r}_{\|}^D(\zeta,y)\bigr)^2}+
\int_{\zeta}^{\infty}y^3dy
\frac{e^{y}\hat{r}_{\bot}^D(\zeta,y)}{\bigl(e^y-
\hat{r}_{\bot}^D(\zeta,y)\bigr)^2}
\biggr]\right\}.
\nonumber
\end{eqnarray}
\noindent
Here, the reflection coefficients for dielectric with the 
frequency-dependent dielectric permittivity are defined in
Eq.~(\ref{eq52}). For many dielectrics, admitting the
presentation (\ref{eq46}) with one oscillator, the characteristic 
frequency at typical separations is much less than the
absorption frequency leading to $b=\omega_c/\omega_1\ll 1$.
In fact the small parameter $b$ is of order of another small
parameter $\eta$. The expansion of Eq.~(\ref{eq52}) in powers
of $b$ takes the form
\begin{eqnarray}
&&
\hat{r}_{\|}^D(x,y)={r}_{\|}^D(x,y)-b^2
\frac{(\varepsilon_0^D-1)x^2y\left[2y^2+
(\varepsilon_0^D-2)x^2\right]}{\sqrt{(\varepsilon_0^D-1)x^2+
y^2}\left(\varepsilon_0^Dy+\sqrt{(\varepsilon_0^D-1)x^2+
y^2}\right)^2}
\nonumber \\
&&\phantom{aaaaaaaaaaaaaaaaaaaaaaaaaaaaaaaaaaaaa}
+\mbox{O}(b^4),
\nonumber \\
&&
\label{eq60} \\
&&
\hat{r}_{\bot}^D(x,y)={r}_{\bot}^D(x,y)-b^2
\frac{(\varepsilon_0^D-1)x^4y}{\sqrt{(\varepsilon_0^D-1)x^2+
y^2}\left(y+\sqrt{(\varepsilon_0^D-1)x^2+
y^2}\right)^2}
\nonumber \\
&&\phantom{aaaaaaaaaaaaaaaaaaaaaaaaaaaaaaaaaaaaa}
+\mbox{O}(b^4),
\nonumber
\end{eqnarray}
\noindent
where $r_{\|,\bot}^D(x,y)$ are the reflection coefficients for
dielectric with a frequency-independent dielectric permittivity
$\varepsilon_0^D$. Our goal is to obtain the expansion of
$\hat{E}(a)$ up to the second powers in the small parameters 
$\eta$ and $b$.

To attain this goal, we note that Eq.~(\ref{eq60}) contains the zeroth 
and second powers in $b$. Thus, both of them should be substituted in 
the zeroth power in $\eta$ in Eq.~(\ref{eq59}). 
Considering the terms of order
$\eta$ and $\eta^2$ in Eq.~(\ref{eq59}) we should restrict ourselves
by only zeroth order in $b$, i.e., replace $\hat{r}_{\|,\bot}(x,y)$
for ${r}_{\|,\bot}(x,y)$. The calculational scheme of all coefficients
accompanying
$\eta$, $\eta^2$ and $b^2$ is the same as was used in Section~3
for obtaining the analytic expression for the function
$\psi_{DM}(\varepsilon_0^D)$. It consists in the expansion of the 
integrands in a power series, changing the order of integrals and
introducing the new variable $w=\zeta/y$.
In the order $\eta^0$ in Eq.~(\ref{eq59}) we obtain the contribution
already calculated in Eqs.~(\ref{eq39}), (\ref{eq40d}) and the
contribution of order $b^2$. The latter takes the form
\begin{eqnarray}
&&
\hat{E}_{b^2}(a)=\frac{\hbar cb^2(\varepsilon_0^D-1)}{32\pi^2a^3}
\sum\limits_{n=1}^{\infty}\int_{0}^{\infty}dyy^4e^{-ny}
\int_{0}^{1}dw\frac{w^2}{\sqrt{(\varepsilon_0^D-1)w^2+1}}
\label{eq61} \\
&&
\phantom{\hat{E}_{b^2}(a)}
\times\left\{\frac{\left[2+(\varepsilon_0^D-
2)w^2\right]\,\left[r_{\|}^D(w)\right]^{n-1}}{\left[\varepsilon_0^D+
\sqrt{(\varepsilon_0^D-1)w^2+1}\right]^2}
+\frac{w^2\left[r_{\bot}^D(w)\right]^{n-1}}{\left[1+
\sqrt{(\varepsilon_0^D-1)w^2+1}\right]^2}
\right\}.
\nonumber 
\end{eqnarray}
\noindent
After the integration with respect to $y$ and summation we arrive at
\begin{eqnarray}
&&
\hat{E}_{b^2}(a)=\frac{3\hbar cb^2}{4\pi^2a^3}
\int_{0}^{1}dw\frac{w^2}{\sqrt{(\varepsilon_0^D-1)w^2+1}}
\label{eq62} \\
&&
\phantom{\hat{E}_{b^2}(a)}
\times\left\{\frac{2+(\varepsilon_0^D-2)w^2}{\varepsilon_0^D+1-w^2}
\mbox{Li}_5\left[r_{\|}^D(w)\right]+
\mbox{Li}_5\left[r_{\bot}^D(w)\right]\right\}.
\nonumber 
\end{eqnarray}

Following the same procedure for the terms of order $\eta$ in 
Eq.~(\ref{eq59}) we obtain
\begin{equation}
\hat{E}_{\eta}(a)=\frac{3\hbar c\eta}{8\pi^2a^3}
\int_{0}^{1}dw
\left\{w^2\mbox{Li}_4\left[r_{\|}^D(w)\right]+
\mbox{Li}_4\left[r_{\bot}^D(w)\right]\right\}.
\label{eq63}
\end{equation}
\noindent
Quite analogically for the terms of order $\eta^2$ in 
Eq.~(\ref{eq59}) it follows
\begin{equation}
\hat{E}_{\eta^2}(a)=-\frac{3\hbar c\eta^2}{2\pi^2a^3}
\int_{0}^{1}dw
\left\{w^4\mbox{Li}_4\left[r_{\|}^D(w)\right]+
\mbox{Li}_4\left[r_{\bot}^D(w)\right]\right\}.
\label{eq64}
\end{equation}

By combining Eqs.~(\ref{eq39}) and (\ref{eq62})--(\ref{eq64})
one arrives at the Casimir energy in the configuration of
metal-dielectric plates made of real materials,
\begin{equation}
\hat{E}(a)=-\frac{\pi^2\hbar c}{720a^3}\psi_{DM}(\varepsilon_0^D)
\left[1-C_1(\varepsilon_0^D)\frac{\delta}{a}+
C_2(\varepsilon_0^D)\frac{\delta^2}{a^2}-
B(\varepsilon_0^D)\frac{\omega_c^2}{\omega_1^2}\right],
\label{eq65}
\end{equation}
\noindent
where the positive coefficients $C_1$, $C_2$ and $B$ are defined as
\begin{eqnarray}
&&
C_1(\varepsilon_0^D)\frac{\delta}{a}\equiv
-\frac{\hat{E}_{\eta}(a)}{E(a)}, \qquad
C_2(\varepsilon_0^D)\frac{\delta^2}{a^2}\equiv
\frac{\hat{E}_{\eta^2}(a)}{E(a)},
\nonumber \\
&&
B(\varepsilon_0^D)\frac{\omega_c^2}{\omega_1^2}\equiv
-\frac{\hat{E}_{b^2}(a)}{E(a)},
\label{eq66}
\end{eqnarray}
\noindent
and $E(a)$ is given in Eq.~(\ref{eq39}).

In Fig.~1 the above coefficients are plotted as functions
of $\varepsilon_0^D$ by the long-dashed lines 1 and 2
($C_1$ and $C_2$, respectively) and by the short-dashed line
($B$). From Eq.~(\ref{eq65}) and Fig.~1 it is easy to obtain the
respective analytic expression for the Casimir pressure,
\begin{equation}
\hat{P}_0(a)=-\frac{\pi^2\hbar c}{240a^4}\psi_{DM}(\varepsilon_0^D)
\left[1-\frac{4}{3}C_1(\varepsilon_0^D)\frac{\delta}{a}+
\frac{5}{3}C_2(\varepsilon_0^D)\frac{\delta^2}{a^2}-
\frac{5}{3}B(\varepsilon_0^D)\frac{\omega_c^2}{\omega_1^2}\right].
\label{eq67}
\end{equation}
\noindent
Equations (\ref{eq65}) and (\ref{eq67}) give the possibility to
simply find the Casimir energy and pressure between metal and
dielectric with rather high precision (see the next section).

From Eqs.~(\ref{eq12}) and (\ref{eq58}) we obtain the asymptotic
behavior of the Casimir entropy at small $\tau$ for
metal-dielectric plates made of real materials,
\begin{eqnarray}
&&
\hat{S}(a,T)=
\frac{3k_B\zeta(3)(\varepsilon_0^D-1)^2}{128\pi^3a^2(\varepsilon_0^D+1)}
\tau^2
\left\{
\vphantom{\left[\frac{\left(\left(\varepsilon_0^D\right)^2
\sqrt{\varepsilon_0^D}\right)}{3(\varepsilon_0^D-1)}\right]}
1-\frac{\pi^2(\varepsilon_0^D+1)}{45\zeta(3)(\varepsilon_0^D-1)}
\tau\right.
\label{eq68} \\
&&
\phantom{\hat{S}}
\times\left.\left[\frac{8\left(1-2\varepsilon_0^D\sqrt{\varepsilon_0^D}+
\left(\varepsilon_0^D\right)^2
\sqrt{\varepsilon_0^D}\right)}{3(\varepsilon_0^D-1)}-(5\varepsilon_0^D+11)\eta
\right]+\mbox{O}(\tau^2)\right\}.
\nonumber
\end{eqnarray}
\noindent
As is seen from Eq.~(\ref{eq68}), $\hat{S}(a,T)$ goes to zero when
the temperature vanishes as is required by the Nernst heat theorem.
This completes the proof of the important statement that the Lifshitz
theory in the configuration metal-dielectric is consistent with
thermodynamics if the static dielectric permittivity of a dielectric
plate is finite.

We complete this section by the consideration of the high-temperature
limit. Here the zero-frequency term (\ref{eq42}) of the Lifshitz
formula determines the total result.  In the configuration of
metal-dielectric plates only the transverse magnetic mode 
(for which the metal reflection coefficient
is equal to unity) contributes
to the zero-frequency term. As a result, unlike the case of two metal plates,
finite conductivity corrections do not contribute at large
separations (high temperatures). Thus, for metal and dielectric
plates made of real materials, equations (\ref{eq43}) and (\ref{eq44})
obtained for ideal metal and dielectric with constant permittivity
preserve their validity.

\section{Comparison between analytic and numerical results}

Here we compare the analytic results for the Casimir energy, free energy
and pressure given by Eqs.~(\ref{eq65}), (\ref{eq67}), (\ref{eq58}) and
(\ref{eq37}) with the results of numerical computations using the
Lifshitz formulas (\ref{eq14}) and the dielectric permittivities
$\varepsilon^{M,D}(i\xi_l)$ determined from the tabulated optical
data for the complex index of refraction. This comparison permits
us to find the applicability regions of the obtained analytic
results for different materials.
As an example we consider the metal plate made of Au and the
dielectric plate made of high-resistivity Si.

The most precise results for $\varepsilon^{M}(i\xi_l)$ in the case of
Au were obtained in \cite{23} and for $\varepsilon^{D}(i\xi_l)$
in the case of Si in \cite{55}. In both cases the data for
Im$\,\varepsilon^{M,D}(\omega)$ were taken from \cite{56} and the 
dielectric permittivities along the imaginary frequency axis were
computed by means of the Kramers-Kronig relation,
\begin{equation}  
\varepsilon^{M,D}(i\xi_l)=1+\frac{2}{\pi}\int_{0}^{\infty}d\omega
\frac{\omega\mbox{Im}\varepsilon^{M,D}(\omega)}{\omega^2+\xi^2}.
\label{eq69}
\end{equation}
\noindent
Note that the dielectric permittivity of Si along the imaginary
frequency axis is equal to its static value
($\varepsilon_0^D=11.66$) up to the angular frequency of
$5\times 10^{14}\,$rad/s and with increase of frequency decreases 
to unity. The analytical results were computed with the plasma
frequency of Au equal to $\omega_p=9.0\,$eV and the characteristic
absorption frequency of Si equal to $\omega_1=4.2\,$eV \cite{56}
($1\,\mbox{eV}=1.519\times 10^{15}\,$rad/s).

In Fig.~2 we compare the results of analytic and numerical
computations of the Casimir energy density (A) and pressure (B)
at different separations at zero temperature. 
In the vertical axes the quantities
$\delta E=(\hat{E}_a-\hat{E}_n)/\hat{E}_n$ (A) and
$\delta P_0=(\hat{P}_{0,a}-\hat{P}_{0,n})/\hat{P}_{0,n}$ (B)
in percent are plotted where $\hat{E}_a$ and $\hat{P}_{0,a}$
are the analytic results calculated by Eqs.~(\ref{eq65}) and
(\ref{eq67}), respectively, and $\hat{E}_n$, $\hat{P}_{0,n}$
are computed numerically using the Lifshitz formula as
described above. As is seen in Fig.~2, the largest deviations 
between the analytic and numerical results (--4.3\% and --7.1\%
for the energy and pressure, respectively) hold at the shortest 
separation of 100\,nm. This is because the plasma model works
good only at separations larger than the plasma wavelength.
At shorter separations not some analytic representations for
$\varepsilon$ but the tabulated optical data should be used to
obtain precise results.
At separations larger than 200 and 300\,nm $|\delta E|$ is
less than 0.9\% and 0.25\%, respectively. As to $|\delta P|$,
it is less than 0.9\% and 0.25\% at respective separations
larger than 250 and 370\,nm. Thus, the obtained analytic
formulas for the Casimir energy density and pressure 
at zero temperature in between 
metal and dielectric give rather precise results in a wide
separation range with a precision at the fraction of a percent.
In some cases this makes unnecessary much more cumbersome numerical
computations using the Lifshitz formula and tabulated optical data
for the complex index of refraction (note that the use of different
sets of tabulated optical data also leads to about 0.5\% differences
in the numerically computed Casimir forces \cite{17}).

In Fig.~3 the results of analytical and numerical computations
of the relative thermal correction to the Casimir energy (A)
and pressure (B) at $a=300\,$nm are compared at different temperatures.
The relative thermal corrections are defined as
\begin{equation}
\frac{\Delta\hat{\mathcal F}(a,T)}{\hat{E}(a)}=
\frac{\hat{\mathcal F}(a,T)-\hat{E}(a)}{\hat{E}(a)},
\quad
\frac{\Delta\hat{P}(a,T)}{\hat{P}_0(a)}=
\frac{\hat{P}(a,T)-\hat{P}_0(a)}{\hat{P}_0(a)},
\label{eq70}
\end{equation}
\noindent
where $\hat{E}(a)$ and $\hat{P}_0(a)$ are the energy density and pressure
calculated numerically by using the Lifshitz formulas at zero temperature.
The analytical computations of the thermal corrections are performed 
using Eqs.~(\ref{eq58}) and (\ref{eq37}). Their results are
shown by the dashed lines. The numerical computations 
of the thermal corrections are done with 
the help of the Lifshitz formula at zero and nonzero temperatures 
(solid lines). As is seen in Fig.~3A, the low-temperature
analytic result for the thermal correction to the
energy density reproduces the result of numerical computations
at $T\leq 20\,$K. From Fig.~3B it follows that the low-temperature
analytic expression for the thermal correction to the
Casimir pressure works good in a wider temperature region
$T\leq 40\,$K. The deviations between analytical and numerical results
at higher temperatures are explained by the fact that in 
Eqs.~(\ref{eq58}) and (\ref{eq37}) we have restrected ourselves by 
only two and one perturbative orders in small parameter $\tau$,
respectively. This restriction, however, makes it possible to solve
the main problem of our interest which has no numerical solution,
i.e., to find the behavior of the Casimir free energy, entropy and
pressure at arbitrarily low temperatures. 

\section{Is the Lifshitz formula for configuration of metal and
dielectric consistent with thermodynamics?}

In Sections~3--5 it was supposed that at zero frequency the dielectric
permittivity of the dielectric plate is finite. It is well known,
however, that at nonzero temperature dielectrics possess some
dc conductivity
$\sigma_0=\sigma_0(T)$ which is very small in comparison with
the conductivity of metals. 
Usually (see, e.g., \cite{44,56a}) this
conductivity is included into the model of dielectric response by
adding a Drude-like term in the dielectric permittivity of
dielectric,
\begin{equation}
\tilde{\varepsilon}^D(i\xi_l)={\varepsilon}^D(i\xi_l)+
\frac{4\pi\sigma_0(T)}{\xi_l}.
\label{eq71}
\end{equation}
\noindent
Eq.~(\ref{eq71}) presents the typical example of the situation
discussed in Section~4 when the dielectric permittivity depends
on the temperature as a parameter. It can be identically represented
in the form
\begin{equation}
\tilde{\varepsilon}^D(i\xi_l)={\varepsilon}^D(i\xi_l)+
\frac{\beta(T)}{l},
\label{eq72}
\end{equation}
\noindent
where $\beta(T)=2\hbar\sigma_0(T)/(k_BT)$. The conductivity of
dielectrics quickly decreases with temperature,
$\sigma_0(T)\sim\exp(-g/T)$, where the coefficient $g$ is
determined by the width of the energy gap $\Delta$ \cite{57}. 
The magnitude of the additional term $\beta(T)/l$ in
Eq.~(\ref{eq72}) is very small. Thus, for SiO${}_2$ at
$T=300\,$K it holds $\beta\sim 10^{-12}$ \cite{58}.
This makes the role of dielectric dc conductivity negligible
at all $l\geq 1$.

The question arises on the possible role of dielectric dc
conductivity in the Casimir interaction between metal and
dielectric. As was shown in \cite{42,43,43a}, the Lifshitz
formula for the Casimir interaction between two dielectrics
cannot incorporate the effects of the dc conductivity because then
an inconsistency with thermodynamics arises.
The substitution of Eq.~(\ref{eq71}) in Eq.~(\ref{eq10}) leads to
\begin{equation}
\tilde{r}_{\|}^D(0,y)=1,
\qquad
\tilde{r}_{\bot}^D(0,y)=0,
\label{eq73}
\end{equation}
\noindent
instead of Eq.~(\ref{eq15}). Despite the negligible role of the dc
conductivity at all $l\geq 1$, this could lead to important
consequences for the Casimir interaction between metal and
dielectric. Importantly, the inclusion of the dc conductivity of
a dielectric plate leads to a discontinuity in the transverse
magnetic reflection coefficient at zero frequency as is seen from
Eqs.~(\ref{eq15}) and (\ref{eq73}). This is unlike the case with
metals described by the Drude model where the discontinuity arises 
in the transverse electric reflection coefficient at zero frequency.

To investigate this problem, we substitute the dielectric permittivity
(\ref{eq72}) in Eq.~(\ref{eq9}) instead of $\varepsilon^D(i\xi_l)$
defined in Eq.~(\ref{eq46}). For a metal, as in Section~4, the dielectric
permittivity in Eq.~(\ref{eq45}) is used. In such a way the Casimir
free energy $\tilde{\mathcal F}(a,T)$ is obtained which takes into account
the effects of the dielectric dc conductivity. It is convenient to
separate the zero-frequency term of $\tilde{\mathcal F}(a,T)$ and
subtract and add the zero-frequency term of the free energy
${\mathcal F}(a,T)$ calculated with the dielectric permittivity
$\varepsilon^D(i\xi_l)$:
\begin{eqnarray}
&&
\tilde{\mathcal F}(a,T)=\frac{k_BT}{16\pi a^2}\int_{0}^{\infty}
ydy\left[\ln\left(1-e^{-y}\right)-\ln\left(1-r_0e^{-y}\right)
\right]
\nonumber \\
&&
\phantom{\tilde{\mathcal F}(a,T)}
+\frac{k_BT}{16\pi a^2}\int_{0}^{\infty}ydy
\ln\left(1-r_0e^{-y}\right)
\label{eq74} \\
&&
\phantom{\tilde{\mathcal F}(a,T)}
+\frac{k_BT}{8\pi a^2}\sum\limits_{l=1}^{\infty}
\int_{\zeta_l}^{\infty}ydy\left\{
\ln\left[1-\hat{r}_{\|}^M(\zeta_l,y)\tilde{r}_{\|}^D(\zeta_l,y)
e^{-y}\right]\right.
\nonumber \\
&&
\phantom{\tilde{\mathcal F}(a,T)}
\left.
+\ln\left[1-\hat{r}_{\bot}^M(\zeta_l,y)\tilde{r}_{\bot}^D(\zeta_l,y)
e^{-y}\right]\right\}.
\nonumber
\end{eqnarray}
\noindent
Here the reflection coefficients $\tilde{r}_{\|,\bot}^D$ are found
by using Eq.~(\ref{eq10}) where the dielectric permittivities
$\varepsilon^D(i\xi_l)$ from Eq.~(\ref{eq46}) are replaced by
$\tilde{\varepsilon}^D(i\xi_l)$ from Eq.~(\ref{eq72}).

To find the behavior of $\tilde{\mathcal F}(a,T)$ at low temperatures,
we expand the last integral on the right-hand side of Eq.~(\ref{eq74})
in powers of the small parameter $\beta(T)/l$. The zero-order contribution 
in this expansion together with the second integral on the right-hand 
side of Eq.~(\ref{eq74}) are equal to the Casimir free energy
$\hat{\mathcal F}(a,T)$ calculated with dielectric permittivity
$\varepsilon^D(i\xi_l)$. Calculating explicitly the first
integral on the right-hand side of Eq.~(\ref{eq74}), we rearrange
this equation to the form
\begin{equation}
\tilde{\mathcal F}(a,T)=\hat{\mathcal F}(a,T)-
\frac{k_BT}{16\pi a^2}\left[\zeta(3)-\mbox{Li}_3(r_0)\right]
+Q(a,T),
\label{eq75}
\end{equation}
\noindent
where $Q(a,T)$ contains all powers in the expansion of the last 
integral on the right-hand side of Eq.~(\ref{eq74})
in the small parameter $\beta(T)/l$ equal or higher than the first one.
The explicit expression for the main, linear in $\beta(T)/l$, term
in $Q(a,T)$ reads:
\begin{eqnarray}
&&
Q_1(a,T)=\frac{k_BT}{8\pi a^2}\sum\limits_{l=1}^{\infty}
\frac{\beta(T)}{l}\int_{\zeta_l}^{\infty}
\frac{dyy^2e^{-y}}{\sqrt{y^2+\zeta_l^2(\varepsilon_l^D-1)}}
\label{eq76} \\
&&
\phantom{Q_1(a,T)}
\times\left\{\frac{(2-\varepsilon_l^D)\zeta_l^2-
2y^2}{\left[\sqrt{y^2+\zeta_l^2(\varepsilon_l^D-1)}+
\varepsilon_l^Dy\right]^2}
\frac{\hat{r}_{\|}^M(\zeta_l,y)}{1-
\hat{r}_{\|}^M(\zeta_l,y)\hat{r}_{\|}^D(\zeta_l,y)e^{-y}}
\right.
\nonumber \\
&&
\phantom{Q_1(a,T)}\left.  
-\frac{\zeta_l^2}{\left[\sqrt{y^2+\zeta_l^2(\varepsilon_l^D-1)}+
y\right]^2}
\frac{\hat{r}_{\bot}^M(\zeta_l,y)}{1-
\hat{r}_{\bot}^M(\zeta_l,y)\hat{r}_{\bot}^D(\zeta_l,y)e^{-y}}
\right\}.
\nonumber
\end{eqnarray}

To determine the asymptotic behavior of Eq.~(\ref{eq76}) when
$\tau\to 0$, we expand the integrated function in powers of
$\tau$ (recall that $\zeta_l=\tau l$) and consider the main
contribution in this expansion at $\tau=0$:
\begin{eqnarray}
&&
Q_1(a,T)=-\frac{k_BTr_0}{4\pi a^2\left(
\left(\varepsilon_0^D\right)^2-1\right)}
\sum\limits_{l=1}^{\infty}
\frac{\beta(T)}{l}\int_{\zeta_l}^{\infty}
\frac{ydye^{-y}}{1-r_0e^{-y}}
\label{eq77} \\
&&
\phantom{Q_1(a,T)}
=-\frac{k_BT\beta(T)}{4\pi a^2\left(
\left(\varepsilon_0^D\right)^2-1\right)}
\sum\limits_{n=1}^{\infty}
\frac{r_0^n}{n^2}\left[\sum\limits_{l=1}^{\infty}
\frac{e^{-n\tau l}}{l}+n\tau\sum\limits_{l=1}^{\infty}
e^{-n\tau l}\right].
\nonumber
\end{eqnarray}
\noindent
Performing the summation in $l$ we obtain
\begin{equation}
Q_1(a,T)=-\frac{k_BT\beta(T)}{4\pi a^2\left(
\left(\varepsilon_0^D\right)^2-1\right)}
\sum\limits_{n=1}^{\infty}
\frac{r_0^n}{n^2}\left[-\ln(1-e^{-n\tau})+
\frac{n\tau}{e^{n\tau}-1}\right].
\label{eq78}
\end{equation}
\noindent
The right-hand side of Eq.~(\ref{eq78}) can be rearranged with
the help of the equality
\begin{equation}
-\ln(1-e^{-n\tau})+
\frac{n\tau}{e^{n\tau}-1}=-\ln\tau+1-\ln{n}+\mbox{O}(\tau^2).
\label{eq79}
\end{equation}
\noindent
As a result it holds
\begin{equation}
Q_1(a,T)=-\frac{k_B\mbox{Li}_2(r_0)}{2\pi a^2\left(
\left(\varepsilon_0^D\right)^2-1\right)}
T\beta(T)\ln\tau+T\beta(T)\mbox{O}(\tau^0).
\label{eq81}
\end{equation}

Taking into account that $\beta(T)\sim(1/T)\exp(-g/T)$, we
arrive at
\begin{equation}
Q_1(a,T)\sim e^{-g/T}\ln{T}.
\label{eq82}
\end{equation}

From Eq.~(\ref{eq82}) it follows that both $Q_1(a,T)$ and its 
derivative with respect to $T$ go to zero when $T$ vanishes.
The terms of higher orders in the small parameters
$\beta(T)/l$ and $\tau$ omitted in our analysis, go to zero
even faster than $Q_1$. Thus, the quantity $Q(a,T)$ in
Eq.~(\ref{eq75}) and its derivative with respect to $T$ have zero
limits when the temperature goes to zero.

Now we are in a position to find the asymptotic behavior of the entropy in
the configuration metal-dielectric with included dc conductivity of the
dielectric plate. Using Eq.~(\ref{eq12}), we obtain from Eq.~(\ref{eq75})
\begin{equation}
\tilde{S}(a,T)=\hat{S}(a,T)+
\frac{k_B}{16\pi a^2}\left[\zeta(3)-\mbox{Li}_3(r_0)\right]
-\frac{\partial Q(a,T)}{\partial T},
\label{eq83}
\end{equation}
\noindent
where $\hat{S}(a,T)$ is defined in Eq.~(\ref{eq68}). In the limit
$T\to 0$ Eq.~(\ref{eq83}) results in
\begin{equation}
\tilde{S}(a,T)=
\frac{k_B}{16\pi a^2}\left[\zeta(3)-\mbox{Li}_3(r_0)\right]>0.
\label{eq84}
\end{equation}
\noindent
This equation implies that in the configuration of metal-dielectric
with included dc conductivity of the dielectric plate the Nernst
heat theorem is violated. Previously the analogous result was
obtained \cite{42,43,43a} for the configuration of two dielectric
plates with frequency-dependent dielectric permittivities. 
It is easily seen, that Eq.~(\ref{eq84}) is preserved, if, instead of 
the plasma dielectric model (\ref{eq45}), the Drude dielectric function 
is used in the case of metal plate with a perfect crystal lattice.
If the metal plate has impurities, the analytical derivation cannot
be performed, but numerical computations lead to the same positive
value of the entropy as in Eq.~(\ref{eq84}).
Thus, the Lifshitz theory becomes inconsistent with
thermodynamics when the dc conductivity of a dielectric plate
is taken into account. This suggests that the actual low-frequency 
behavior of the dielectric properties is not related to the phenomena of
van der Waals and Casimir forces and should not be included
into the model of dielectric response.

Recently \cite{25c} this theoretical conclusion was confirmed
experimentally in the measurement of the difference Casimir force
acting between Au-coated sphere and Si plate illuminated by laser
pulses. The difference of the Casimir forces in the presence and in
the absence of pulse was measured using an atomic force microscope.
In the absence of laser pulse the concentration of charge
carriers was of about $5\times 10^{14}\,\mbox{cm}^{-3}$
(higher-resistivity Si), but in the presence of pulse this
concentration has been enhanced up to $2\times 10^{19}\,\mbox{cm}^{-3}$
(lower-resistivity Si). The experimental data were compared with
two theories. The first theory used an assumption that in the
absence of laser light Si possesses a finite static dielectric 
permittivity $\varepsilon_0^{D}=11.66$ (see Section~5).
The second theory took into account the dc conductivity of Si in the
absence of laser light like it was done above in Section~6. 
The first theory was found to be in excellent agreement with data,
whereas the second theory was excluded at the 95\% confidence level 
 within the separation region from 100 to 200\,nm \cite{25c}.
Thus, the inclusion of the dc conductivity of dielectrics and
high-resistivity semiconductors in the model of dielectric
response is not only inconsistent thermodynamically but is
also in contradiction with experiment.

\section{Conclusions and discussion}

In this paper we have obtained the analytic expressions for the Casimir free 
energy, pressure and entropy at low temperatures in the configuration of
one metal and one dielectric plate. Different models of the dielectric
response for both metal and dielectric were considered: the simplified
model of an ideal metal and dielectric with constant dielectric
permittivity, and the realistic model of a metal described by the plasma
model and a dielectric described in the Ninham-Parsegian representation
for $\varepsilon$. To derive the asymptotic expressions at low temperatures,
the perturbation theory in the small parameter $\tau$ was developed which is
proportional to the product of separation distance and the temperature.
The analytic expressions for the main physical quantities in the limit of
high temperatures and at zero temperature were also obtained. The
analytic results were compared with numerical computations using the 
Lifshitz formula and tabulated optical data, and good agreement was found. 

The fundamental conclusion arrived in the paper is that the Lifshitz
theory applied to the configuration of one metal and one dielectric
plate is in agreement with thermodynamics if the dielectric permittivity
of a dielectric plate at zero frequency is finite. In particular, it was
shown that the Casimir entropy goes to zero when the temperature
vanishes, i.e., the Nernst heat theorem is satisfied. 
This conclusion cannot be reached numerically, because it is impossible to
perform numerical computations at arbitrarily low temperature 
and their precision is 
always restricted.
On the contrary, it was shown that,
if the dielectric permittivity of the dielectric plate at zero frequency
turns to infinity (i.e., small dc conductivity of a dielectric material
is taken into account), this leads to a nonzero value of
the Casimir entropy at zero temperature, i.e., to a violation of the
Nernst heat theorem. 
The inclusion of the dc conductivity of high-resistivity
semiconductors in the model of dielectric response was also recently
shown to be in contradiction with experiment \cite{25c}.
What this means is that to avoid contradictions
with thermodynamics and experiment,
 one should not include the actual conductivity 
properties of dielectric materials at very low, quasistatic, frequencies
into the model of the dielectric response (this phenomenological
prescription was obtained
previously in \cite{42,43,43a} for the case of two dielectric plates
made of real materials).

The above conclusions can be discussed in the context of the
formalism of thermal quantum field theory in Matsubara formulation
where the zero-frequency term plays a separated role and calls for
an adequate interpretation. It is common knowledge that the zero-point
energy of quantized fields contains oscillations of any frequency.
It would be hard, however, to imagine the presence of a zero-frequency
(i.e., constant) field in the vacuum state. This suggests that the
zero-frequency term in the Matsubara summation could be understood
not literally but as a mathematical limit to zero from the region of
much higher frequencies [the characteristic frequency of the Casimir
effect $c/(2a)$, and the thermal frequency $k_{B} T/\hbar$] which determine
the physical phenomena of the van der Waals and Casimir forces. As was
discussed below Eq.~(\ref{eq72}), in the region of characteristic and thermal
frequencies the effects of dc conductivity contribute twelve orders of
magnitude less than $\varepsilon^D$. Because of this, the dc conductivity
may be considered
as not related to the Casimir forces and be not included into
the zero-frequency term of the Lifshitz formula.
This phenomenological prescription is not the fundamental resolution
of the problem, which remains unknown, but following it one avoids
contradictions to thermodynamics and experiment. 

Note that the problems discussed above for the configuration of
metal-dielectric are of different nature than those arising for two 
metals. For metals the concentration of charge carriers only
slightly depends on the temperature. The validity of the Nernst
heat theorem in the Casimir interaction between two metals is caused
by the scattering processes of free charge carriers on phonons,
impurities etc. For the perfect Drude metals with no impurities, 
relaxation goes to zero when the temperature vanishes and the Nernst heat
theorem is violated \cite{30,40,41}. On the contrary, for dielectrics
the concentration of charge carriers quickly decreases to zero when
the temperature vanishes. Here the violation of the Nernst heat
theorem does not depend on the scattering processes and is caused by the
inclusion of the infinitely large dc conductivity. 
In the formalism, for two metals a discontinuity in the reflection
coefficient of the transverse electric mode at zero frequency arises,
whereas for a metal and dielectric a discontinuity holds in the
reflection coefficient of the transverse magnetic mode.
These differences are
reflected in the fact that even the sign of the entropy at $T=0$ for two
perfect Drude metals and for one metal and one dielectric with included
dc conductivity are opposite (negative for two metals and positive for 
a metal and dielectric).

The obtained results are topical for the interpretation of recent 
measurements of the Casimir force between metal sphere and semiconductor
plate \cite{25,25a,25b,25c}. Semiconductors suggest a wide variety of the
conductivity properties ranging from metallic to dielectric ones. In the
application of the Lifshitz theory to the metal-semiconductor test bodies 
the model of the dielectric response for a semiconductor should be
chosen to satisfy the Nernst heat theorem and other fundamental
physical principles. This can be done by using the proposed phenomenological
prescription. A more fundamental resolution of the discussed problems may 
go beyond the scope of the Lifshitz theory.

\section*{Acknowledgments}

G.~L.~K. and V.~M.~M. are grateful to the Center of Theoretical
Studies and the Institute for Theoretical Physics, Leipzig University for
their kind hospitality. This work was supported by Deutsche 
Forschungsgemeinschaft, Grant No. 436 RUS 113/789/0-2.


\newpage
\begin{figure}[h]
\vspace*{-7cm}
\includegraphics{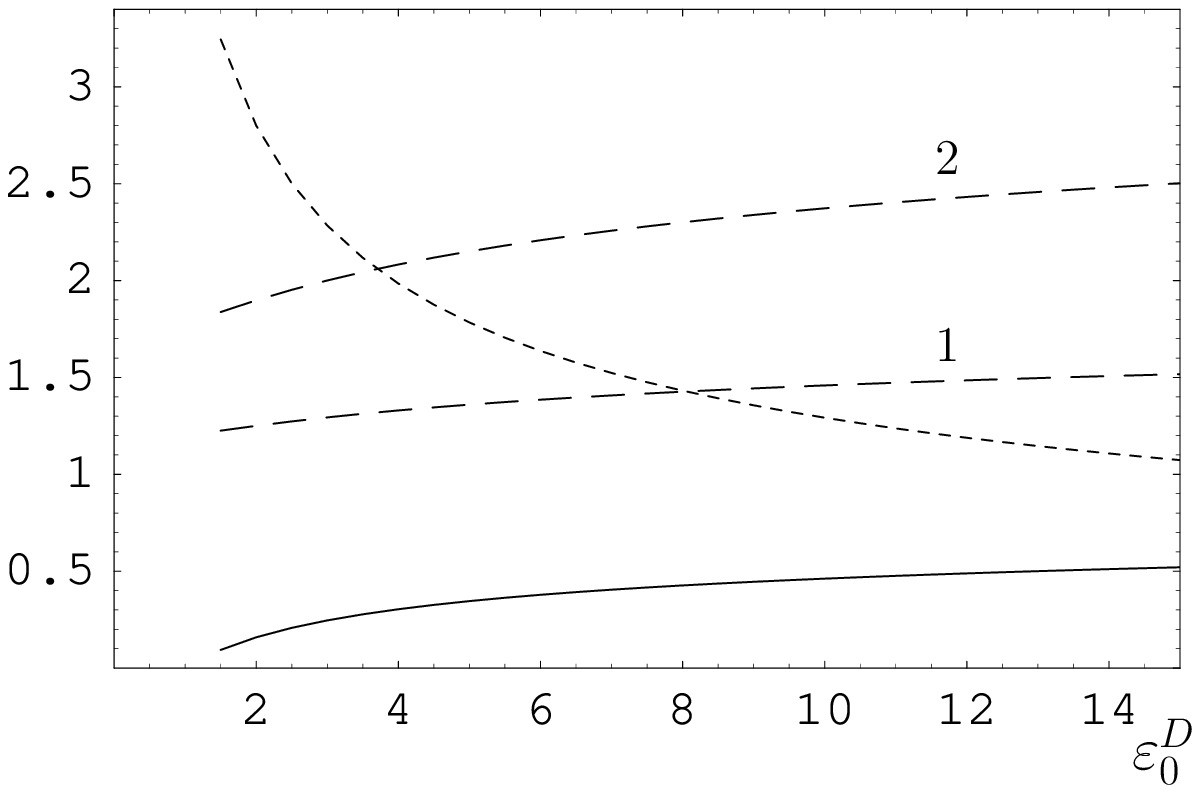}
\vspace*{-10cm}
\caption{The correction factor $\psi_{DM}$ to the Casimir energy
density (solid line) as a function of the static dielectric
permittivity. The long-dashed lines\ 1,\ 2 and the short-dashed line
show the coefficients $C_1$, $C_2$ and $B$, respectively,
in Eq.~(\ref{eq65}) for the Casimir energy density between plates made
of real metal and dielectric. 
}
\end{figure}
\begin{figure}[h]
\vspace*{-2cm}
\includegraphics{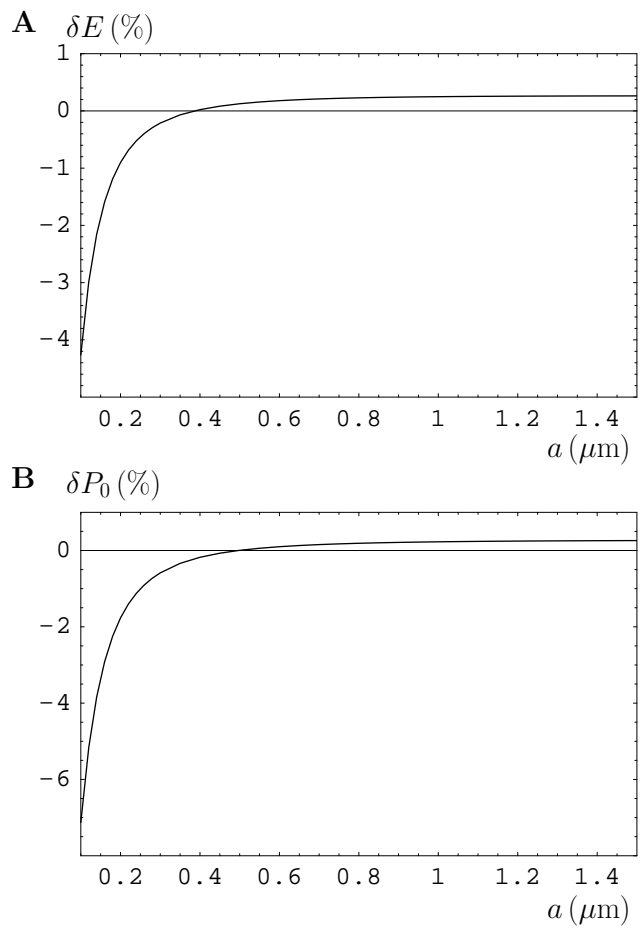}
\vspace*{-11cm}
\caption{The relative differences between the analytic and numerical
results for the Casimir energy density (A) and pressure (B)
at zero temperature versus separation.
}
\end{figure}
\begin{figure}[h]
\vspace*{-2cm}
\includegraphics{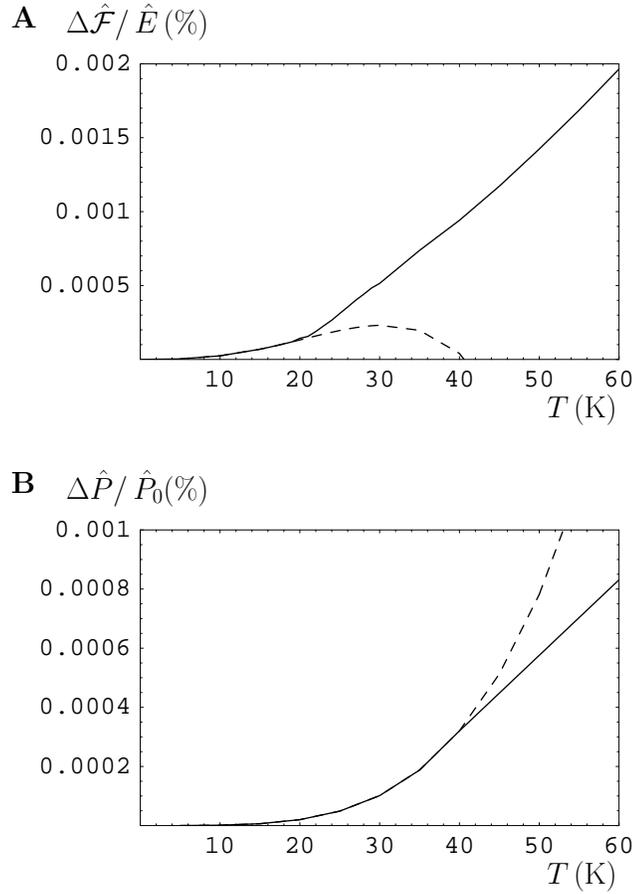}
\vspace*{-11cm}
\caption{The relative thermal correction to the Casimir energy density (A) 
and pressure (B) versus temperature at a separation $a=300\,$nm. 
Solid lines show the results
of numerical computations and the dashed lines are obtained 
using the analytic asymptotic expressions.
}
\end{figure}

\end{document}